\documentclass[aps,prb,twocolumn,showpacs,floatfix,superscriptaddress]{revtex4}
\pdfoutput=1

\usepackage[english]{babel}
\usepackage[pdftex]{graphicx}
\usepackage{float}
\usepackage{hyperref}
\hypersetup{colorlinks,linkcolor=blue,urlcolor=blue,citecolor=blue}
\usepackage{amsfonts,amssymb,amsmath}

\newcommand{\Hi}{\hat{H}_{\rm i}}
\newcommand{\Hf}{\hat{H}_{\rm f}}
\newcommand{\Gi}{\ket{G_{\rm i}}}
\newcommand{\Gf}{\ket{G_{\rm f}}}
\newcommand{\ginitial}{G_{\rm i}}
\newcommand{\gfinal}{G_{\rm f}}
\newcommand{\Vl}{V_{\rm large}}
\newcommand{\GXt}{\mathcal{G}_{X} (t)}
\newcommand{\GYt}{\mathcal{G}_{Y} (t)}
\newcommand{\AXw}{\mathcal{A}_{X} ( \omega )}
\newcommand{\G}{\mathcal{G}}
\newcommand{\A}{\mathcal{A}}
\newcommand{\ostar}{\omega^{\ast}}
\newcommand{\barostar}{\bar \omega^{\ast}}
\newcommand{\omegahe}{\omega_{\rm he}}
\newcommand{\teqzero}{t = 0}
\newcommand{\othresh}{\omega_{\rm th}}
\newcommand{\NC}{n_{\mathrm{c}}}

\newcommand{\equil}{{\rm eq}}
\newcommand{\atan}{\tan^{-1}}
\newcommand{\initial}{{\rm i}}
\newcommand{\final}{{\rm f}}
\newcommand{\tot}{{\rm tot}}
\newcommand{\loc}{{\rm dot}}
\newcommand{\sea}{{\rm sea}}
\newcommand{\bath}{{\rm B}}
\newcommand{\charge}{{\rm ch}}
\newcommand{\Gren}{\Gamma_{\rm ren}}
\newcommand{\GammaL}{\Gamma_{\!\rm L}}
\newcommand{\GammaR}{\Gamma_{\!\rm R}}
\newcommand{\Gammamu}{\Gamma_{\!\!\mu}}
\newcommand{\DPS}{W_{\rm PS}}
\newcommand{\DPSS}{W_{\rm PS}^{\rm S}}
\newcommand{\US}{U_{\rm S}}
\newcommand{\Beff}{B_{\rm eff}}
\newcommand{\TK}{T_{\rm K}}

\newcommand{\TKS}{T_{\rm K}^{\rm S}}
\newcommand{\epsd}{\varepsilon_{\!d}}
\newcommand{\pdag}{{\phantom{\dagger}}}
\newcommand{\ce}{\hat{c}_{\varepsilon}^{\pdag}}
\newcommand{\ced}{\hat{c}_{\varepsilon}^{\dagger}}
\newcommand{\cem}{\hat{c}_{\varepsilon \mu}^{\pdag}}
\newcommand{\cemd}{\hat{c}_{\varepsilon \mu}^{\dagger}}
\newcommand{\kF}{k_{\rm F}}
\newcommand{\dt}{\Delta_{\rm AO}}
\newcommand{\dmulti}{\Delta_{{\rm AO} , \mu}}
\newcommand{\dch}{\Delta_{\rm ch}}
\newcommand{\dtX}{\Delta_{\! X}}
\newcommand{\dtXsq}{\dtX^{2}}
\newcommand{\dtYsq}{\dtY^{2}}
\newcommand{\dtC}{\Delta_{\!\!\; C}}
\newcommand{\dtY}{\Delta_{\!\!\; Y}}
\newcommand{\dtd}{\Delta_{d}}
\newcommand{\dtdph}{\Delta_{d}^{\rm ph}}
\newcommand{\dth}{\Delta_{\!\!\; h}}
\newcommand{\dtm}{\Delta_{\mu}}
\newcommand{\dtL}{\Delta_{\rm L}}
\newcommand{\dtR}{\Delta_{\rm R}}
\newcommand{\dtS}{\Delta_{\rm S}}
\newcommand{\dX}{\eta_{X}}

\newcommand{\dCequil}{\eta_{C}^\equil}
\newcommand{\dd}{\eta_{d}}
\newcommand{\ddp}{\eta_{d}^0}
\newcommand{\ddc}{\eta_{dc}}
\newcommand{\ddcp}{\eta_{dc}^0}
\newcommand{\ddct}{\eta_{dc^{\dagger}}}
\newcommand{\ddctp}{\eta^0_{dc^\dagger}}
\newcommand{\etap}{\eta^0}
\newcommand{\dhd}{\eta_{hd}}
\newcommand{\dhdc}{\eta_{hdc}}
\newcommand{\dhdct}{\eta_{hdc^{\dagger}}}
\newcommand{\dYS}{\eta_{Y}^{\rm S}}
\newcommand{\dY}{\eta_{Y}}
\newcommand{\dYp}{\eta_{Y}^0}
\newcommand{\ket}[1]{\vert #1 \rangle }
\newcommand{\bra}[1]{\langle #1 \vert}
\newcommand{\braket}[2]{\langle #1 \vert #2 \rangle}
\newcommand{\qph}{\quad \phantom{.}}
\newcommand{\qqph}{\qquad \phantom{.}}

\newcommand{\qm}[1]{``#1''}
\newcommand{\komma}{\:,}
\newcommand{\punkt}{\:.}
\newcommand{\Eq}[1]{Eq.~(\ref{#1})}
\newcommand{\Eqs}[1]{Eqs.~(\ref{#1})}
\newcommand{\eref}[1]{Eq.~(\ref{#1})}
\newcommand{\Eref}[1]{Equation~(\ref{#1})}
\newcommand{\erefs}[1]{Eqs.~(\ref{#1})}
\newcommand{\Erefs}[1]{Equations~(\ref{#1})}
\newcommand{\erefb}[1]{(\ref{#1})}
\newcommand{\sref}[1]{Sec.~\ref{#1}}
\newcommand{\Sref}[1]{Section~\ref{#1}}
\newcommand{\fref}[1]{Fig.~\ref{#1}}
\newcommand{\Fref}[1]{Figure~\ref{#1}}
\newcommand{\frefs}[1]{Figs.~\ref{#1}}

\newcommand{\frefsub}[2]{Fig.~\ref{#1}(#2)}
\newcommand{\Frefsub}[2]{Figure~\ref{#1}(#2)}
\newcommand{\frefsuba}[3]{Figs.~\ref{#1}(#2) and (#3)}
\newcommand{\frefsubt}[3]{Figs.~\ref{#1}(#2-#3)}
\newcommand{\PR}{\textit{Phys. Rev. }}
\newcommand{\PRB}{\textit{Phys. Rev.} B }
\newcommand{\PRL}{\textit{Phys. Rev. Lett. }}
\newcommand{\RMP}{\textit{Rev. Mod. Phys. }}
\newcommand{\EPJ}{\textit{Eur. Phys. J.} B }

\begin{document}

\title{Anderson Orthogonality in the Dynamics After a Local Quantum Quench}	

\author{Wolfgang \surname{M\"under}}
\affiliation{
	Physics Department, Arnold Sommerfeld Center for Theoretical Physics, and Center for NanoScience, \\
	Ludwig-Maximilians-Universit\"at, Theresienstrasse 37, 80333 Munich, Germany
}
\author{Andreas \surname{Weichselbaum}}
\affiliation{
	Physics Department, Arnold Sommerfeld Center for Theoretical Physics, and Center for NanoScience, \\
	Ludwig-Maximilians-Universit\"at, Theresienstrasse 37, 80333 Munich, Germany
}
\author{Moshe \surname{Goldstein}}
\affiliation{
	Department of Physics, Yale University, 217 Prospect Street, New Haven, Connecticut 06520, USA
}
\author{Yuval \surname{Gefen}}
\affiliation{
	Department of Condensed Matter Physics, The Weizmann Institute of
	Science, Rehovot 76100, Israel
}
\author{Jan \surname{von Delft}}
\affiliation{
	Physics Department, Arnold Sommerfeld Center for Theoretical Physics, and Center for NanoScience, \\
	Ludwig-Maximilians-Universit\"at, Theresienstrasse 37, 80333 Munich, Germany
}

\begin{abstract}
We present a systematic study of the role of Anderson orthogonality 
for the dynamics after a quantum quench in quantum impurity models, 
using the numerical renormalization group. As shown by Anderson in 
1967, the scattering phase shifts of the single-particle wave functions 
constituting the Fermi sea have to adjust in response to the sudden 
change in the local parameters of the Hamiltonian, causing the initial 
and final ground states to be orthogonal. This so-called Anderson 
orthogonality catastrophe also influences dynamical properties, such 
as spectral functions. Their low-frequency behaviour shows nontrivial 
power laws, with exponents that can be understood using a generalization 
of simple arguments introduced by Hopfield and others for the X-ray edge 
singularity problem. The goal of this work is to formulate these generalized 
rules, as well as to numerically illustrate them for quantum quenches in 
impurity models involving local interactions. As a simple yet instructive 
example, we use the interacting resonant level model as testing ground 
for our generalized Hopfield rule. We then analyse a model exhibiting 
population switching between two dot levels as a function of gate voltage, 
probed by a local Coulomb interaction with an additional lead serving as 
charge sensor. We confirm a recent prediction that charge sensing can 
induce a quantum phase transition for this system, causing the population 
switch to become abrupt. We elucidate the role of Anderson orthogonality 
for this effect by explicitly calculating the relevant orthogonality exponents.
\end{abstract}

\date{\today}


\maketitle

%---------------------------------------------------------------------
\section{Introduction}\label{sec:Intro}
%---------------------------------------------------------------------

The Anderson orthogonality (AO) catastrophe \cite{Anderson1967} refers
to the response of a Fermi sea to a change in a local scattering
potential, described, say, by a change in Hamiltonian from $\Hi$ to
$\Hf$. Such a change induces changes in the scattering phase shifts of
all single-particle wave functions.  This causes the initial ground
state $\ket{\ginitial}$ of $\Hi$ and the final ground state
$\ket{\gfinal}$ of $\Hf$, both describing a filled Fermi sea but
w.r.t. different single-particle wave functions, to be orthogonal in
the thermodynamic limit, even if the changes in the single-particle
wave functions are minute. The overlap of the respective ground states
scales as \cite{Anderson1967,Schotte1969}
\begin{equation}
	\vert \braket{\ginitial}{\gfinal} \vert \sim N^{- \frac{1}{2} \dt^2}
	\komma \label{eq:Intro_overlap}
\end{equation}
where $N$ is the number of particles in the system, and the exponent $\dt$ 
characterizes the degree of orthogonality.

AO underlies the physics of numerous dynamical phenomena such as the
Fermi edge singularity, \cite{Mahan1967,Schotte1969,Hopfield1969,Nozieres} 
the Altshuler-Aronov zero bias anomaly \cite{Altshuler1979} in disordered
conductors, tunnelling into strongly interacting Luttinger liquids,
\cite{Gogolin1993,Prokofev1994,Affleck1994,Furusaki1997,Gogolin} slow 
relaxation in electron glasses, \cite{Leggett1987,Ovadyahu2007} and 
optical absorption involving a Kondo exciton,\cite{Helmes2005,Tureci2011,Latta2011} 
where photon absorption induces a local quantum quench, to name but a few.
Recently, AO has also been evoked \cite{Goldstein2010,Goldstein2011}
in an analysis of \emph{population switching} (PS) in quantum dots
(the fact that the population of individual levels of a quantum dot
may vary non-monotonically with the gate voltage), and was argued to
lead, under certain conditions involving a local Coulomb interaction
with a nearby charge sensor, to a quantum phase transition.

One of the goals of the present work is to analyse the latter
prediction in quantitative detail. Another is to generalize arguments
that were given in Refs.~\onlinecite{Helmes2005,Tureci2011,Latta2011},
for the role of AO for spectral functions of the excitonic Anderson
model, to related models with a similar structure. Thus, we present a
systematic study of the role of Anderson orthogonality for the
dynamics after a quantum quench in quantum impurity models involving
local interactions, using the numerical renormalization group
(NRG).\cite{Wilson1975,Bulla2008} We thereby extend a recent study, 
\cite{Weichselbaum2011} which showed how $\dt$ can be calculated very
accurately (with errors below $1\%$) by using NRG to directly evaluate
overlaps such as  $\braket{\ginitial}{\gfinal}$, to the domain of
dynamical quantities.

The spectral functions that characterize a local quantum quench
typically show power-law behaviour, ${\sim \omega^{-1 + 2 \eta}}$, in
the limit of small frequencies, where $\eta$ typically depends on
$\dt$.\cite{Mahan1967,Schotte1969,Hopfield1969,Nozieres} For the case
of the X-ray edge singularity, Hopfield \cite{Hopfield1969} gave a
simple argument to explain the relation between $\dt$ and $\eta$.  We
frame Hopfield's argument in a more general setting and numerically
illustrate the validity of the resulting generalized Hopfield rule
(\Eq{eq:Ay} below) for several nontrivial models. In particular, we 
also analyse how this power-law behaviour is modified at low 
frequencies when one adds to the Hamiltonian an extra tunnelling term,
that describes transitions between the Hilbert spaces characterizing
the \qm{initial} and \qm{final} configurations. This effect plays a
crucial role in understanding the abovementioned quantum phase
transition for population switching.

The paper is organized as follows. In \sref{sec:AOconsequences} we
review various consequences of AO in different but related settings,
and formulate the abovementioned generalization of Hopfield's rule.
In \sref{sec:IRLM} we illustrate this rule for the spinless
interacting resonant level model (IRLM), involving a single localized
level interacting with the Fermi sea of a single lead. We consider
this model without and with tunnelling, and study a quantum quench of
the energy of its local level, focussing on signatures of AO in each
case. Finally, in \sref{sec:PS_A} and \sref{sec:PS_C} we discuss
population switching without and with a charge sensor, respectively,
confirming that if the  sensor is sufficiently strongly coupled,
AO indeed does cause population switching to become
a sharp quantum phase transition.
\Sref{sec:Conclusions} offers concluding remarks and outlines
prospective applications of the present analysis.

%---------------------------------------------------------------------
\section{Various Consequences of Anderson Orthogonality} \label{sec:AOconsequences}
%---------------------------------------------------------------------

In this section we review various consequences of AO, in different 
but related settings. We begin by recalling two well-known facts: 
first, the relation between the exponent $\dt$ and the charge that 
is displaced due to the quantum quench, $\dch$; and second, the 
role of $\dt$ in determining the asymptotic long-time power-law 
decay of correlation functions $\G_{X} (t)$ involving an operator 
$\hat{X}^{\dagger}$ that connects the initial and final ground state. 

Then we consider the spectral function $\A_{X} (\omega)$ associated 
with $\G_{X} (t)$, which correspondingly shows asymptotic power-law 
behaviour, ${\sim \omega^{-1 + 2 \eta}}$, for small frequencies, where 
the exponent $\eta$ depends on $\dt$. We recall and generalize an 
argument due to Hopfield, that extends the relation between $\eta$ 
and $\dt$ to composite local operators. Finally, we recapitulate how 
all these quantities can be calculated using NRG.

For simplicity, we assume in most of this section that the Fermi sea 
consists only of a single species of (spinless) electrons.The generalization 
to several channels needed in subsequent sections (in particular for 
discussing PS), is straightforward and will be introduced later as needed. 

Although the concepts summarized in subsections~\ref{sec:displaced-charge} 
to \ref{sec:NRG} below apply quite generically to a wide range of impurity models, 
for definiteness we will illustrate them by referring to a particularly simple example, 
to be called the \qm{local charge model} (LCM), which we define next. 

%---------------------------------------------------------------------
\subsection{Local charge model}\label{sec:localchargemodel}
%---------------------------------------------------------------------

\begin{figure}
\begin{centering}
\includegraphics{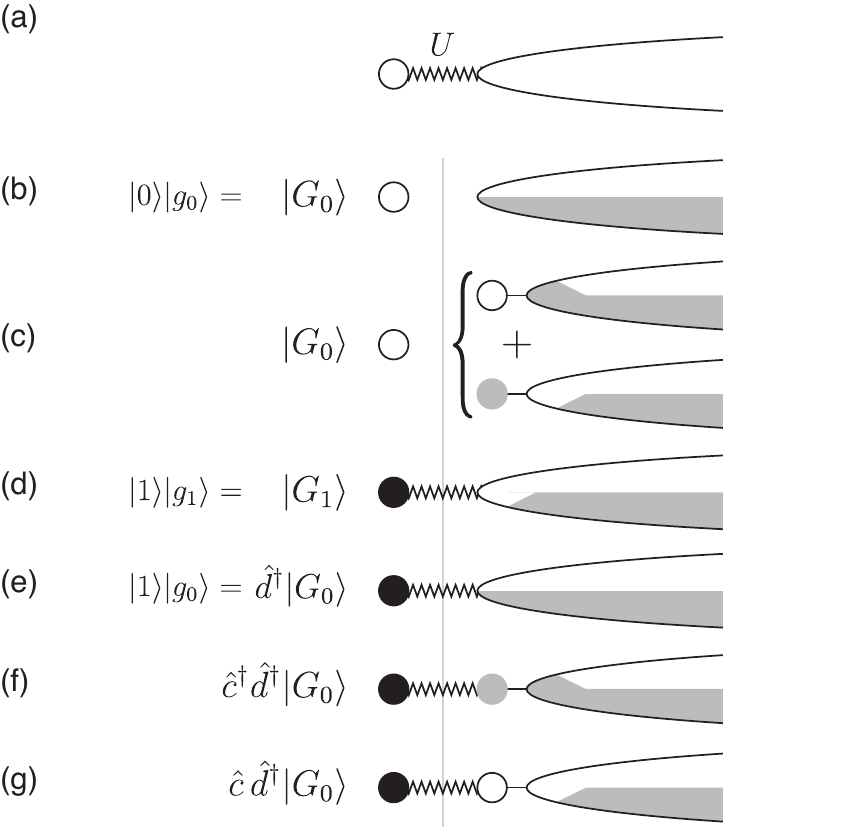}
\par
\end{centering}
\caption{\label{fig:IRLM_A}
(a) Cartoon of the Hamiltonian~\erefb{eq:LCM} for the LCM. 
(b) to (g) Cartoons of the occupation of the dot and a half-filled lead, 
for $U>0$, for  several states discussed in the text. (b) and (c) give two 
equivalent depictions of the ground state $\ket{G_{0}}$ of $\hat{H}_{0}$. 
(c) depicts the fact that $\ket{g_{0}}$ can be written as a superposition 
of the form $\ket{0}_{\rm c} \ket{Q}_{\rm rest} + \ket{1}_{\rm c} \ket{Q-1}_{\rm rest}$, 
indicating complementary occupations of the first site and the rest of a 
half-filled Wilson chain (defined in \sref{sec:NRG} below). Here $\ket{0}_{\rm c}$ 
(which obeys $\hat{c} \ket{0}_{\rm c} = 0$) and $\ket{1}_{\rm c} = \hat{c}^{\dagger} 
\ket{0}_{\rm c}$ describe the first site of the Wilson chain being empty or 
filled, respectively; the charge in the rest of the Wilson chain is correspondingly 
distributed in such a way that both components of the superposition have the 
same total charge, $Q$. (d) depicts the ground state $\ket{G_{1}}$ of $\hat{H}_{1}$, 
indicating that charge on the dot pushes charge in the lead away from the dot site. 
(e) shows the effect of applying $\hat{d}^{\dagger}$ to $\ket{G_{0}}$, the latter 
depicted according to (b). Similarly, (f) and (g) show the effect of applying 
$\hat{c}^{\dagger} \hat{d}^{\dagger}$ or $\hat{c} \hat{d}^{\dagger}$ to 
$\ket{G_{0}}$, the latter depicted according to (c). The displaced charge 
flowing inwards from infinity towards the dot as each of the states (e) to 
(g) evolves to the final ground state $\ket{G_{1}}$ of (d) is $\dtd < 0$, 
$\dtd - 1 < 0$ or $\dtd + 1 > 0$, respectively. Comparison of (f) and 
(g) with (e) shows average charge differences of $+1$ and $-1$, respectively, 
in accord with the Hopfield-type argument summarized by \eref{eq:deltaC}.
}
\end{figure}

The LCM describes a single spinless localized level, to be called dot 
level (alluding to a localized level in a quantum dot), interacting with 
a single Fermi sea of spinless electrons [see \frefsub{fig:IRLM_A}{a}]:
\begin{equation}
	\hat{H}_{\rm LCM} (\hat{n}_{d}) =
	U \, \hat{n}_{d} \, \hat{c}^{\dagger} \hat{c}
	+ \sum_{\varepsilon} \varepsilon \, \ced \ce
	\punkt \label{eq:LCM}
\end{equation}
Here $\ce$ and $\hat{d}$ are annihilation operators for Fermi sea states 
and the dot state, respectively, $\hat{n}_{d} =\hat{d}^{\dagger} \hat{d}$ 
counts the number of dot electrons, and $\hat{c} \equiv \hat{\psi} (0) 
\equiv \sum_{\varepsilon} \ce$ destroys a Fermi sea electron at the 
position of the dot. The interaction is taken to be repulsive, $U>0$. There 
is no tunnelling between dot and sea. Therefore, the Hilbert space separates 
into two distinct sectors, in which the local charge operator $\hat{n}_{d}$ 
has eigenvalues $n_{d} = 0$ and $n_{d} = 1$, respectively. The Hamiltonians 
describing the Fermi sea in the two distinct sectors are
\begin{subequations}
\begin{eqnarray}
  \hat{H}_{0} & = & \hat H_{\rm LCM} (n_d = 0)
  = \sum_{\varepsilon} \varepsilon \, \ced \ce
  \komma \label{eq:IRLM_H0} \\
  \hat{H}_{1} (U) & = & \hat H_{\rm LCM} (n_d = 1)
  = \sum_{\varepsilon} \varepsilon \, \ced \ce
  + U \hat{c}^{\dagger} \hat{c}
  \punkt \label{eq:IRLM_H1} \qqph
\end{eqnarray}
\end{subequations}
We will denote their respective ground states [illustrated in 
Figs.~\ref{fig:IRLM_A}(b,c) and \ref{fig:IRLM_A}(d), respectively] by 
\begin{eqnarray}
	\ket{G_{0}}  = \ket{0} \ket{g_{0}}
	\komma \label{eq:SIRL_A_G0} \qquad
	\ket{G_{1}}  = \ket{1} \ket{g_{1}}
	\komma \label{eq:SIRL_A_G1}
\end{eqnarray}
where $\ket{0}$ and $\ket{1} = \hat{d}^{\dagger} \ket{0}$ describe 
the dot state with charge 0 or 1, respectively, and $\ket{g_{0}}$ and 
$\ket{g_{1}}$ the corresponding Fermi sea ground states. 

The LCM contains all ingredients needed for AO, hence we will repeatedly 
refer to it below as an explicit example of the general arguments to be 
presented. [Corresponding LCM passages will sometimes appear in square 
brackets, so as not to disrupt the general flow of the discussion.]  Explicit 
numerical results for the LCM will be presented in \sref{sec:IRLM_A} below. 

%---------------------------------------------------------------------
\subsection{AO and the displaced charge} \label{sec:displaced-charge}
%---------------------------------------------------------------------

For the ensuing discussions, it will be useful to distinguish between 
two types of quenches, to be called type 1 and 2, which we now discuss 
in turn.

\emph{Type 1 quench}: For a type 1 quench, some parameter of the 
Hamiltonian is changed abruptly (e.g.\ by a sudden change of gate 
voltage for one of the gates defining a quantum dot). Taking the LCM 
as an example, suppose that the value of the interaction in the LCM is 
changed suddenly from $U$ to $U'$ for a \emph{fixed} local charge of 
$n_{d} = 1$. This corresponds to a type 1 quench with
\begin{subequations}
\label{eq:GiGfLCMtype1}
\begin{align}
	\Hi & = \hat{H}_{1} (U) \, , \quad
	& \Hf & = \hat{H}_{1} (U') \, ,
	\label{eq:GiGfLCMtype1-Hamiltonians} \\
	\ket{G_{\initial}} & = \ket{1} \ket{g_{1,\initial}} \, , \quad
	& \ket{G_{\final}} & = \ket{1} \ket{g_{1,\final}} \, . \phantom{(}
	\label{eq:GiGfLCMtype1-states}
\end{align}
\end{subequations}
The overlap of initial and final ground states, 
\begin{equation}
	\vert \braket{G_{\initial}}{G_{\final}} \vert = 
	\vert \braket{g_{1,\initial}}{g_{1,\final}} \vert 
	\sim N^{-\frac{1}{2} \dt^2}
	\komma \label{eq:GiGftype1}
\end{equation}
will vanish in the thermodynamic limit due to AO, since the two 
Fermi sea states $\ket{g_{1,\initial}}$ and $\ket{g_{1,\final}}$ feel
scattering potentials of different strengths.

In his classic 1967 paper, Anderson showed that for this type of
situation the exponent $\dt$ in \Eq{eq:GiGftype1} is equal to the
change in scattering phase shifts at the Fermi surface divided by 
$\pi$, in reaction to the change in the strength of the scattering
potential. According to the Friedel sum rule,
\cite{Friedel1952,Friedel1956,Langreth1966, Hewson} the change in
phase shifts divided by $\pi$, in turn, is equal to the
\emph{displaced charge} $\dch$ (in units of $e$) that flows inward
from infinity into a large but finite volume (say $\Vl$) surrounding
the scattering site, in reaction to the change in scattering
potential, so that $\dt = \dch $.  To be explicit,
\begin{eqnarray}
	 \dch \equiv
	\bra{G_{\final}} \hat{n}_\tot \ket{G_{\final}}
	- \bra{G_{\initial}} \hat{n}_\tot \ket{G_{\initial}} \; ,
	\label{eq:Intro_Friedel}
\end{eqnarray}
where $\hat n_\tot \equiv \hat n_\sea + n_\loc$ counts the 
\emph{total} number of electrons within $\Vl$, with $\hat{n}_{\sea}$ 
counting the Fermi sea electrons and $\hat{n}_{\loc}$ counting the 
electrons on the dot. [For the LCM, $\hat{n}_{\loc} = \hat{n}_{d}$.] 

The relative sign between $\dt$ and $\dch$ ($+$ not $-$) is a matter 
of convention, which does not affect the orthogonality exponent $\dt^2$. 
Our convention,\cite{Weichselbaum2011} which agrees with standard 
usage, \cite{StandardUsage} is such that $\dt > 0$ (or $<0$) if the change 
in local potential induces electrons to flow inward toward (outward away 
from) the scattering site. 

For the LCM quench of \eref{eq:GiGfLCMtype1} above, the initial and 
final states have the same dot charge, $n_{d} = 1$, hence the displaced 
charge reduces to $\dch \equiv \bra{g_{1,\final}} \hat{n}_{\sea} \ket{g_{1,\final}} 
- \bra{g_{1,\initial}} \hat{n}_{\sea} \ket{g_{1,\initial}}$. However, such a 
simplification will not occur for more complex impurity models involving 
tunneling between dot and lead [of the form $( \hat{d}^{\dagger} \hat{c} 
+ \hat{c}^{\dagger} \hat{d} )$], so that the local charge is not conserved. 
Examples are the interacting resonant level model [\eref{eq:SIRL_B_H} below], 
or the single-impurity Anderson model [\eref{eq:SIAM} below]. 

For such a model, consider a type 1 quench from $\Hi$ to $\Hf$, implemented 
by a sudden change in one or several model parameters, in analogy to \eref{eq:GiGfLCMtype1-Hamiltonians}. 
Although the corresponding ground states $\ket{G_{\initial}}$ and $\ket{G_{\final}}$ 
will no longer have the simple factorized form of \eref{eq:GiGfLCMtype1-states}, they 
will still exhibit AO as in \eref{eq:Intro_overlap}. Moreover, the decay exponent is still 
equal  to the displaced charge, $\dt = \dch$, given by \eref{eq:Intro_Friedel}. (For 
a NRG verification of this fact, see Ref.~\onlinecite{Weichselbaum2011}.) 

\emph{Type 2 quench}: For a type 2 quench, all model parameters 
are kept constant, but the system is switched suddenly between two 
dynamically disconnected sectors of Hilbert space characterized by 
different conserved quantum numbers. Taking again the LCM of 
\eref{eq:LCM} as an example, suppose that the local charge is suddenly 
changed, say from $n_{d} = 0$ to $1$, while all model parameters are 
kept  constant. This corresponds to a type 2 quench with 
\begin{subequations}
\label{eq:GiGfLCMtype2}
\begin{align}
	\Hi & = \hat{H}_{0} \phantom{U} \, \quad
	& \Hf & = \hat{H}_{1} \; , \phantom{U} \qph
	\label{eq:GiGfLCMtype2-Hamiltonians} \\
	\Gi & = \ket{0} \ket{g_{0}} , \quad
	& \Gf & = \ket{1}	 \ket{g_{1}} . \qph
	\label{eq:GiGfLCMtype2-states}
\end{align}
\end{subequations}
A physical example of such a quench would be core level X-ray 
photoemission spectroscopy (XPS), where an incident X-ray photon 
is absorbed by an atom in a crystal, accompanied by the ejection of 
a core electron from the material. \cite{Gunnarsson1983a} This 
amounts to the sudden creation of a core hole, which subsequently 
interacts with the Fermi sea of mobile conduction electrons (but does 
not hybridize with them). Thus, in this example $\hat{n}_{d}$ would 
represent the hole number operator $\hat{n}_{h} = \hat{h}^{\dagger} 
\hat{h}$.

More generally, a type 2 quench presupposes a Hamiltonian $\hat{H} 
(\hat{n}_{x})$ that depends on a conserved charge, say $\hat{n}_{x}$ 
[such as $\hat{n}_{d}$ for the LCM], with eigenvalues $n_{x}$ [such 
as $n_{d} = 0$ or $1$]. The Hilbert space can then be decomposed 
into distinct, dynamically disconnected sectors, labelled by $n_{x}$ 
and governed by effective Hamiltonians $\hat{H} (n_{x})$, whose 
ground states have the form $\ket{G(n_x)} = \ket{n_{x}} \ket{g(n_x)}$. 
A type 2 quench is induced by an operator, say $\hat{X}^{\dagger}$ 
[such as $\hat{d}^{\dagger}$ for the LCM], whose action changes the 
conserved charge, thereby connecting two distinct sectors, say $\bra{n'_{x}} 
\hat{X}^{\dagger} \ket{n_{x}} = 1$, with $n'_{x} \neq n_{x}$. For such 
a quench we make the identifications
\begin{subequations}
\label{eq:GiGfLCMtype2-nx}
\begin{align}
	\Hi & = \hat{H} (n_{x}) \, \quad
	& \Hf & = \hat{H} (n'_{x}) \, , \phantom{\ket{n_{x}}} \qph
	\label{eq:GiGfLCMtype2-Hamiltonians-nx} \\
	\Gi & = \ket{n_{x}} \ket{g (n_{x})} , \quad
	& \Gf & = \ket{n_{x}'} \ket{g (n_{x}')} . \qph
	\label{eq:GiGfLCMtype2-states-nx}
\end{align}
\end{subequations}
The overlap $\braket{G_{\initial}} {G_{\final}} = 0$ vanishes trivially, 
because $\braket{n_{x}}{n_{x}'} = 0$.  However, define 
\begin{eqnarray}
	\ket{\psi_{\initial}} \equiv \hat{X}^{\dagger} \Gi
	\label{eq:post-quench-psi_i}
\end{eqnarray}
to be the \qm{initial post-quench state} obtained by the action of 
the charge switching operator $\hat{X}^{\dagger}$ on the initial 
ground state. [\frefsub{fig:IRLM_A}{e} illustrates this state for the 
LCM with $\hat{X}^{\dagger} = \hat{d}^{\dagger}$.] Then the overlap 
\begin{equation}
	\mathcal{O}_{X} \equiv \vert \braket{\psi_{\initial}}{G_{\final}} \vert
	= \vert \braket{g(n_{x})}{g(n_{x}')} \vert
	\sim N^{-\frac{1}{2} \dtXsq}
	\label{eq:SIRL_A_Od} 
\end{equation}
again shows AO, since it is equal to the overlap of two Fermi sea ground 
states corresponding to different local charges. The corresponding exponent 
in \eref{eq:SIRL_A_Od} can again be related to a displaced charge, $\dtX = 
\dtX^{\charge}$, but now the latter should compare the total charge within 
$\Vl$ described by the states $\Gf$ and $\ket{\psi_{\initial}}$:
\begin{eqnarray}
	 \dtX^{\charge} \equiv \bra{G_{\final}} \hat{n}_{\tot} \ket{G_{\final}}
	- \bra{\psi_{\initial}} \hat{n}_\tot \ket{\psi_{\initial}}
	\punkt \label{eq:Intro_Friedel-X}
\end{eqnarray}
$\dtX^{\charge}$ can be interpreted as the charge (in units of $e$) that flows 
into $\Vl$ during the post-quench time evolution from $\ket{\psi_{\initial}}$ 
to $\Gf$ subsequent to the action of $\hat{X}^{\dagger}$. To simplify notation, 
we will often omit the superscript $\charge$ distinguishing the displaced charge 
$\dtX^{\charge}$ from the AO exponent $\dtX$, since the two are equal in any case. 

\emph{Composite type 2 quench}: Let us now consider a more complicated 
version of a type 2 quench, induced by a composite operator of the form 
$\hat{Y}^{\dagger} = \hat{C}^{\dagger} \hat{X}^{\dagger}$. Here $\hat X^\dagger$ 
switches between disconnected sectors of Hilbert space as above, while 
$\hat{C}^{\dagger}$ does not; instead, $\hat{C}^{\dagger}$ is assumed 
to be a \emph{local} operator which acts on the dot or in the Fermi sea at 
the location of the dot, but commutes with $\hat{n}_{x}$. For the LCM, an 
example would be $\hat{C}^{\dagger} = \hat{c}^{\dagger}$, so that $\hat{Y}^{\dagger}$ 
creates two electrons, one on the dot, one in the Fermi sea at the site of the 
dot. 

A physical realization hereof is furnished by the edge-ray edge effect 
occurring in X-ray absorption spectroscopy (XAS), where an incident X-ray 
photon is absorbed by an atom in a crystal, accompanied by the creation of 
a core hole ($\hat{X}^{\dagger} = \hat{h}^{\dagger}$) and the transfer of a 
core electron into the conduction band of the metal ($\hat{C}^{\dagger} = 
\hat{c}^{\dagger}$). \cite{Gunnarsson1983a} Another example is the Kondo 
exciton discussed in Refs.~\onlinecite{Tureci2011,Latta2011}, where the 
absorption of a photon by a quantum dot is accompanied by the creation 
of an electron-hole pair on the dot, described by $\hat{C}^{\dagger} = 
\hat{e}^{\dagger}$ and $\hat{X}^{\dagger} = \hat{h}^{\dagger}$, respectively. 
In this example, the hole number $\hat{n}_{h} = \hat{h}^{\dagger} \hat{h}$ 
is conserved, but the dot electron number $\hat{n}_{e} = \hat{e}^{\dagger} 
\hat{e}$ is not, since the Hamiltonian contains dot-lead hybridization terms 
of the form $(\hat{e}^{\dagger} \hat{c} + \hat{c}^{\dagger} \hat{e})$ (see 
Refs.~\onlinecite{Tureci2011,Latta2011} for details). 

For a composite type 2 quench, the initial and final 
Hamiltonians and ground states are defined as in \Eqs{eq:GiGfLCMtype2-nx},
but the post-quench initial state is given by 
\begin{eqnarray}
	\ket{\psi_{\initial}'} \equiv \hat{Y}^{\dagger} \Gi = 
	\hat{C}^{\dagger} \ket{\psi_{\initial}}
	\punkt \label{eq:post-quench-psi_i-Y}
\end{eqnarray}
Its overlap with the final ground state $\ket{G_{\final}'}$ 
to which it evolves in the long time limit has the form
\begin{equation}
	\mathcal{O}_{Y} \equiv 
	\vert \braket{\psi_{\initial}'}{\gfinal'} \vert =
	\vert \bra{g (n_{x}^{\phantom{,}})} \hat{C} \ket{g' (n_{x}')} \vert
	\sim N^{-\frac{1}{2} \dtYsq}
	\punkt \label{eq:SIRL_A_Od-Y}
\end{equation}
The exponent $\dtY$ arising here is related to $\dtX$ and can be found 
using the following argument, due to Hopfield. \cite{Hopfield1969} Due 
to the action of $\hat{C}^{\dagger}$, the states $\ket{\psi_{\initial}'}$ and 
$\ket{\psi_{\initial}}$ describe different amounts of initial post-quench 
charge within the volume $\Vl$. We will denote the difference by
\begin{equation}
	\dtC \equiv
	\bra{\psi_{\initial}'} \hat{n}_{\tot} \ket{\psi_{\initial}'}
	- \bra{\psi_{\initial}} \hat{n}_{\tot} \ket{\psi_{\initial}}
	\punkt \label{eq:deltaC}
\end{equation}

For example, if $\hat{C}^{\dagger}$ is a local electron creation or 
annihilation operator, then $\dtC = 1$ or $-1$, respectively [as 
illustrated in \frefsuba{fig:IRLM_A}{f}{g}]. However, since an initial 
charge surplus or deficit at the scattering site is compensated, in 
the long-time limit, by charges flowing to or from infinity, the ground 
states $\ket{\gfinal'}$ and $\Gf$ towards which $\ket{\psi_{\initial}'}$ 
and $\ket{\psi_{\initial}}$ evolve, respectively, will differ only by one 
Fermi sea electron at infinity, and hence for practical purposes describe 
the same local physics. In particular, the charge within $\Vl$ is the 
same for both, $\bra{\gfinal'} \hat{n}_{\tot} \ket{\gfinal'} = \bra{\gfinal} 
\hat{n}_{\tot} \Gf$. Therefore, the \emph{total} displaced charge 
associated with the action of $\hat{Y}^{\dagger}$ is 
\begin{equation}
	\dtY \equiv \bra{\gfinal'} \hat{n}_{\tot} \ket{\gfinal'}
	- \bra{\psi_{\initial}'} \hat{n}_{\tot} \ket{\psi_{\initial}'}
	= \dtX - \dtC ,
	 \label{eq:related-displaced-charges}
\end{equation}
where the second equality follows from \erefs{eq:deltaC} and 
\erefb{eq:Intro_Friedel-X}. The exponent governing the AO decay 
in \eref{eq:SIRL_A_Od-Y} is thus given by \eref{eq:related-displaced-charges}. 
Since $\dtC$  is a trivally known integer, knowledge of $\dtX$ for 
a type 2 quench suffices to determine the AO exponents $\dtY$ 
for an entire family of related composite quenches. 
 
To conclude this section, we note that a type 1 quench can always 
be formulated as a type 2 quench, by introducing an auxiliary 
conserved degree of freedom (say $\hat{n}_{h}$), whose only purpose 
is to divide the Hilbert space into two sectors (labelled by $n_{h} = 0$ 
or $1$), within which some parameters of the Hamiltonian take two 
different values. For example, if the quench involves changing $U$ 
to $U'$, this can be modelled by replacing $U$ by $U + \hat{n}_{h} 
(U'-U)$ in the Hamiltonian. For an example, see \sref{sec:IRLM_C}. 

%---------------------------------------------------------------------
\subsection{AO and post-quench time evolution}\label{sec:AO-Gt}
%---------------------------------------------------------------------

After a sudden change in the local Hamiltonian, AO also affects 
the long-time limit of the subsequent time evolution, and hence 
the low-frequency behaviour of corresponding spectral functions. 
A prominent example is optical absorption, \cite{Mahan1967,Schotte1969,
Hopfield1969,Nozieres,Helmes2005,Tureci2011,Latta2011} for which 
AO leaves its imprint in the shape of the absorption spectrum, by 
reducing the probability for absorption. This is familiar from the x-ray 
edge problem. \cite{Mahan1967} In particular, in the limit of absorption 
frequency $\omega$ very close to (but above) the threshold for absorption, 
the zero-temperature absorption spectrum has a power-law form, with an 
exponent that is influenced by AO.  Recent demonstrations of this fact can 
be found in studies, both theoretical \cite{Helmes2005,Tureci2011} and 
experimental, \cite{Latta2011} of exciton creation in quantum dots via 
optical absorption, whereby an electron is excited from a valence-band 
level to a conduction band level.

In this subsection, we will analyse the role of AO for the time evolution 
after a type 2 quench of the form \erefb{eq:GiGfLCMtype2}. We consider 
the following generic situation: For $t<0$, a system is in the ground 
state $\Gi$ of the initial Hamiltonian $\Hi$ (with ground state energy 
$E_{\initial}$), describing a Fermi sea under the influence of a local 
scattering potential. At $\teqzero$, a sudden change in the local potential 
occurs, described by the action the local operator $\hat{X}^{\dagger}$. 
It switches sector $n_{x}$ to $n_{x}'$, yielding the post-quench initial 
state $\ket{\psi_{\initial}} = \hat{X}^{\dagger} \Gi$ at time $\teqzero^+$, 
and switches the Hamiltonian from $\Hi$ to $\Hf$. 

The subsequent  dynamics can be characterized by the correlator
\begin{equation}
	\GXt \equiv -i e^{i \omega_{0} t} \theta (t)
	\bra{G_{\initial}} \hat{X} ( t ) \hat{X}^{\dagger} \ket{G_{\initial}}
	\komma \label{eq:defineGX}
\end{equation}
where $\hat{X} ( t ) = e^{i \Hi t} \hat{X} e^{-i \Hf t}$, reflecting
the fact that $\hat{X}$ switches $\Hf$ to $\Hi$.  The phase factor
$e^{i \omega_{0} t}$ is included for later convenience, with
$\omega_{0}$ to be specified below [after \Eq{subeq:SpectralFunction}]. 

Since the Fermi sea adjusts in reaction to the sudden change in local 
potential at $\teqzero$, AO builds up and the overlap function $\G_{X} 
( t )$ decreases with time.  It is known since 1969 that in the long-time 
limit it decays in power-law fashion as \cite{Schotte1969,Hopfield1969}
\begin{eqnarray}
	\GXt \sim t^{- \dtXsq}
	\komma \label{eq:Intro_Correlator2}
\end{eqnarray}
where $\dtX$ is the exponent governing the AO decay of $\mathcal{O}_X$ 
in \eref{eq:SIRL_A_Od}. This can be understood heuristically by expanding 
\eref{eq:defineGX} as 
\begin{subequations}
\begin{eqnarray}
	i e^{-i (E_{\initial} + \omega_{0}) t} \GXt
	& = & \theta ( t ) \bra{\psi_{\initial}} e^{-i \Hf t} \ket{\psi_{\initial}}
	\label{eq:Intro_GX2} \\
	& = & \theta ( t ) \braket{\psi_{\initial}}{\psi_{\initial} ( t )} \\
	& = & \theta ( t ) \sum_{n} e^{-i E_{n} t} 
\vert \braket{\psi_{\initial}}{n} \vert^{2}
	\komma \label{eq:GXn} \qqph
\end{eqnarray}
\end{subequations}
where $\ket{\psi_{\initial} ( t )} = e^{-i \Hf t}
\ket{\psi_{\initial}}$ describes the time-evolution for $t > 0$, and 
$\ket{n}$ and $E_{n}$ represent a complete set of eigenstates and 
eigenenergies of $\Hf$. In the long-time limit \eref{eq:GXn} will be 
dominated by the ground state $\Gf$ of $\Hf$ (with eigenenergy 
$E_{\final}$), yielding a contribution $\vert \braket{\psi_{\initial}}{\gfinal} 
\vert^2$ that scales as $N^{-{\dtXsq}}$ [by \eref{eq:SIRL_A_Od}]. Now, 
as time increases, the effect of the local change in scattering potential 
is felt at increasing length scales $L(t) \sim v_{f} t$, with $v_{f}$ the Fermi 
velocity; regarding $\Gf$ as the lowest eigenstate of $\Hf$ in a box of 
size $N \sim L(t)$, the AO of $\vert \braket{\psi_{\initial}}{\gfinal} \vert^2 
\sim L(t)^{-\dtXsq}$ implies \eref{eq:Intro_Correlator2}. 

For a composite type 2 quench induced by $\hat{Y}^{\dagger} = 
\hat{C}^{\dagger} \hat{X}^{\dagger}$, we can conclude by analogous 
arguments that 
\begin{eqnarray}
	\GYt \sim t^{- \dtYsq}
	\komma \label{eq:Intro_Correlator2-Y}
\end{eqnarray}
where $\dtY$ is the displaced charge of \eref{eq:related-displaced-charges}. 

For future reference, we also introduce the correlator 
\begin{eqnarray}
	\G_{C}^{\equil} (t) \equiv -i \theta(t)
	\bra{G} e^{i \hat{H} t} \hat{C} e^{-i \hat{H} t} \hat{C}^{\dagger} \ket{G}
	\sim i t^{- 2 \dCequil}
	\label{eq:define-scaling-dimension}
\end{eqnarray}
of an operator $\hat{C}^\dagger$ that does not switch between dynamically 
disconnected sectors, i.e.\ that commutes with $\hat{n}_{x}$ [examples of 
such operators are given in the discussion before \eref{eq:post-quench-psi_i-Y} 
above]. Then \eref{eq:define-scaling-dimension} is a standard \emph{equilibrium} 
correlator, with $\Hi = \Hf$, in contrast to the \emph{quench} correlator $\GXt$ 
of \eref{eq:define-scaling-dimension}, where $\Hi \neq \Hf$. For such an 
equilibrium correlator the decay exponent $\dCequil$ is called the scaling 
dimension of $\hat{C}^{\dagger}$. A local operator $\hat{C}^{\dagger}$ is 
relevant, marginal or irrelevant under renormalization if $\dCequil < 1$, 
$=1$ or $>1$, respectively. \cite{Cardy1996} 

%---------------------------------------------------------------------
\subsection{AO and spectral functions}\label{sec:AO-spectral-functions}
%---------------------------------------------------------------------

Next we  consider the spectral function corresponding to $\GXt$, 
\begin{subequations}
\label{subeq:SpectralFunction}
\begin{eqnarray}
	\AXw & \equiv &
	-\frac{1}{\pi} \Im \left( \intop_{0}^{\infty} \mathrm{dt}
	e^{i (\omega + i0^{+} ) t} \GXt \right)
	\label{eq:defineA} \\
	& = & \sum_{n} \vert \bra{n} \hat{X}^{\dagger} \ket{G_{\rm i}} \vert^{2}
	\delta ( \omega - E_{n} +  E_{\initial} + \omega_{0} )
	\punkt \qquad
\end{eqnarray}
\end{subequations}
It  evidently has the form of a golden-rule transition rate for 
$\hat{X}^{\dagger}$-induced transitions with excitation energy 
$\omega + \omega_{0}$ and is nonzero only for $\omega$ 
above the threshold frequency $\othresh = ( E_\final
- E_{\initial} ) - \omega_{0}$. For simplicity, we will here and 
henceforth set $\othresh = 0$ by choosing $\omega_{0} = 
E_\final - E_{\initial}$. Note the sum rule $\int d\omega 
\A (\omega) = \bra{G_\initial} \hat{X} \hat{X}^\dagger \ket{G_\initial}$, 
which can be used as consistency check for numerical calculations.

\Eref{eq:Intro_Correlator2} implies that in the limit $\omega 
\to \othresh = 0$, the spectral function behaves as
\begin{equation}
	\A_{X} ( \omega ) \sim \omega^{-1 + 2\dX}
	\komma \quad
	\dX = \tfrac{1}{2} \dtXsq 
	\punkt \label{eq:Intro_A_X} 
\end{equation}
The definition of $\dX$ is deliberately chosen such that \eref{eq:Intro_A_X} 
parallels the form of the equilibrium spectral function corresponding 
to $\G_{C}^{\equil} (t)$ of \eref{eq:define-scaling-dimension}, namely 
\begin{eqnarray}
	\A_{C}^{\equil} (\omega) \sim \omega^{-1 + 2 \dCequil} \punkt
	\label{eq:Aequil-scaling-dimension}
\end{eqnarray}

Now consider the spectral function $\A_{Y} (\omega)$ involving 
the composite type 2 quench operator $\hat{Y}^{\dagger} = 
\hat{C}^{\dagger} \hat{X}^{\dagger}$. \Erefs{eq:Intro_Correlator2-Y} 
and \erefb{eq:related-displaced-charges} immediately lead to the prediction
\begin{eqnarray}
	\A_{Y} ( \omega ) & \sim & \omega^{-1 + 2\dY}
	\komma \quad \label{eq:Ay}
	\dY  =  \tfrac{1}{2} (\dtX - \dtC)^2
	\komma \label{eq:Hopfield_dY} \qqph
\end{eqnarray}
to be called the \emph{generalized Hopfield rule}, since the essence 
of the argument by which we have obtained it was first formulated by 
Hopfield. \cite{Hopfield1969}

A physical situation for which \eref{eq:Hopfield_dY} is relevant is the 
edge-ray edge effect occurring in X-ray absorption spectroscopy (XAS). 
There we have $\hat{Y}^{\dagger} = \hat{c}^{\dagger} \hat{h}^{\dagger}$ 
(as explained above), and $\dtC = 1$. Thus \eref{eq:Hopfield_dY} yields 
\begin{equation}
	\A_{h c} ( \omega ) \sim \omega^{-1 + (\dth - 1)^2}
	= \omega^{-2\dth + \dth^2}
\label{eq:A_X-ray}
\end{equation}
reproducing a well-established result for the X-ray edge absorption spectrum 
[Ref.~\onlinecite{Hopfield1969}, p.~48; Ref.~\onlinecite{Nozieres}, Eq.~(66)].
In the literature, $-2 \dth$ is often called the \qm{Mahan contribution} to the 
exponent, and $\dth^2$ the AO contribution. Since $\dth \le 1$, one has 
$2 \dth > \dth^2$, i.e.\ \qm{Mahan wins}, and $\A_{h c} ( \omega )$ diverges 
at small frequencies. For present purposes, though, it is perhaps somewhat 
more enlightening to adopt  Hopfield's point of view, stated in \eref{eq:Ay}, 
according to which both terms, $-2\dth$ and $\dth^2$ arise from the AO 
exponent $(\dth - 1)^2$.

\Erefs{eq:SIRL_A_Od}, (\ref{eq:Intro_A_X}) and \erefb{eq:Hopfield_dY} will 
play a central role in this work. Their message is that the near-threshold 
behaviour of spectral functions of the type defined in \eref{subeq:SpectralFunction} 
is governed by an AO exponent that can be extracted from the overlap 
$\braket{\psi_{\initial}}{G_{\final}}$ between the initial post-quench 
state  $|\psi_\initial \rangle$ and the ground state $\ket{G_\final}$ 
to which it evolves in the long-time limit.

To conclude this section, we remark that the above analysis generalizes 
straightforwardly to models involving several species or channels of 
electrons, say with index $\mu$, provided that the channel index 
is a conserved quantum number (i.e.\ no tunnelling between channels 
occurs). \cite{Weichselbaum2011} Then the initial and final ground 
states will be  products of the ground states for each separate channel, 
so that \eref{eq:Intro_overlap} generalizes to 
\begin{eqnarray}
	\vert \braket{\ginitial}{\gfinal} \vert \sim \prod_{\mu}
	N_{\mu}^{- \frac{1}{2} \dmulti^2}
	\punkt \label{eq:multichannel}
\end{eqnarray}
All power laws discussed above that involve $\dt^2$ (or quantities
derived therefrom) in the exponent can be similarly generalized by 
including appropriate products over channels. 

%---------------------------------------------------------------------
\subsection{AO exponents and NRG}\label{sec:NRG}
%---------------------------------------------------------------------

Results of the above type have been established analytically, in the
pioneering papers from 1969, Refs.~\onlinecite{Nozieres,Schotte1969,
Mahan1967,Hopfield1969}, only for the simple yet paradigmatic case 
of the X-ray edge effect. Nevertheless, \eref{eq:Ay} can be expected to 
hold for a larger class of models, as long as the setting outlined above 
applies. Indeed, it has recently been found to hold also in the context 
of the Kondo exciton. \cite{Helmes2005,Tureci2011,Latta2011} The 
purpose of this work, therefore, is to establish the validity of the 
connections between \erefs{eq:SIRL_A_Od}, \erefb{eq:Intro_A_X} 
and \erefb{eq:Hopfield_dY} for a series of models of increasing
complexity. We shall do so numerically using NRG, since for most 
of these models an analytical treatment along the lines of 
Refs.~\onlinecite{Schotte1969} and \onlinecite{Nozieres} would 
be exceedingly tedious, if not impossible. However, the requisite 
numerical tools are available within NRG, \cite{Oliveira_Cox,Costi1994} 
and have become very accurate quantitatively due to recent 
methodological refinements. \cite{Tureci2011,Anders2005,Anders2006}

NRG, developed in the context of quantum impurity models, offers a 
very direct way of evaluating the overlap, since it allows both ground 
states $\Gi$ and $\Gf$ to be calculated explicitly. Models treatable by 
NRG have the generic form $\hat{H} = \hat{H}_{\bath} + \hat{H}_{d}$. 
Here 
\begin{equation}
	\hat{H}_{\bath} = \sum_{\mu = 1}^{\NC} \sum_{\varepsilon}
	\varepsilon \, \cemd \cem
	\komma \label{eq:freeFermiSeas}
\end{equation}
describes a free Fermi sea involving $\NC$ channels of fermions, with 
constant density of states $\rho$ per channel and half-bandwidth $D 
= 1/(2\rho)$. (When representing numerical results, energies will be 
measured in units of half-bandwidth by setting $D=1$.) $\hat{H}_{d}$, 
which may involve interactions, describes local degrees of freedom and 
their coupling to the Fermi sea.

Wilson discretized the spectrum of $\hat{H}_{0}$ on a logarithmic grid
of energies $\pm D \Lambda^{-k}$ (with $\Lambda > 1$, $k = 0, 1, 2,
\dots$), thereby obtaining exponentially high resolution of low-energy
excitations. He then mapped the impurity model onto a semi-infinite
\qm{Wilson tight-binding chain} of sites $k= 0$ to $\infty$, with the
impurity degrees of freedom coupled only to site $0$. To this end, he
made a basis transformation from the set of Fermi sea operators $\{
\cem \}$ to a new set $\{ \hat{f}_{k \mu} \}$, with $\hat{f}_{0 \mu} 
\propto \hat{c}_{\mu} \equiv \psi_{\mu} (0) \equiv \sum_{\varepsilon} 
\cem$, chosen such that they bring $\hat{H}_{0}$ into the tridiagonal 
form 
\begin{equation}
	\hat{H}_\bath \simeq \sum_{\mu = 1}^{\NC}  \sum_{k=1}^{\infty}
	t_{k} ( \hat{f}_{k \mu}^{\dagger} \hat{f}_{k-1 , \mu}^{\pdag} + {\rm h.c.} )
	\komma \label{eq:WilsonChainH0}
\end{equation}
with hopping matrix elements $t_{k} \propto D \Lambda^{-k/2}$ 
that decrease exponentially with site index $k$ along the chain. 
Because of this separation of energy scales, the Hamiltonian can 
be diagonalized iteratively by solving a Wilson chain of length $k$ 
(restricting the sum in \Eq{eq:WilsonChainH0} to the first $k$ terms) 
and increasing $k$ one site at a time. The number of kept states at 
each iteration will be denoted by $N_{k}$. 

For a Wilson chain of length $k$, the effective level spacing of its 
lowest-lying energy levels is set by the smallest hopping matrix 
element of the chain, namely $\Lambda^{-k/2}$; such a Wilson 
chain thus represents a real space system of volume $\Vl \sim 
\Lambda^{k/2}$. Thus, the overlap between the two ground states 
of a Wilson chain of length $k$ can be expressed as \cite{Weichselbaum2011}
\begin{equation}
	\vert {}_{k} \braket{\ginitial}{\gfinal}_{k} \vert
	\sim \Lambda^{- \frac{k}{4} \dt^{2}}
	\equiv e^{- \alpha k}
	\komma \label{eq:Intro_NRG}
\end{equation}
where $\alpha \equiv (\log\Lambda/4) \dt^{2}$.  Explicit calculations 
show \cite{Weichselbaum2011} that an exponential decay of the form 
\eref{eq:Intro_NRG} applies for the overlap between any two states 
$\ket{E_{\initial}}_{k}$ and $\ket{E_{\final}}_{k}$ representing low-lying 
excitations w.r.t.\ $\Gi_{k}$ and $\Gf_{k}$ at iteration $k$, respectively. 
More technically, ${}_{k} \braket{E_{\initial}}{E_{\final}}_{k} \sim e^{- \alpha k}$ 
holds whenever $\ket{E_{\initial}}_{k}$ and $\ket{E_{\final}}_{k}$ represent 
NRG eigenstates with matching quantum numbers from the $k$-th NRG 
shell for $\Hi$ and $\Hf$, respectively, and their overlap is calculated for 
increasing $k$. For multi-chain models, we note that channel-specific 
exponents such as $\dmulti$ [see \Eq{eq:multichannel}] can be calculated, 
if needed, by considering Wilson chains with channel-dependent lengths. 
\cite{Weichselbaum2011} 

Within the framework of NRG, a consistency check is available for the
value of $\dt$ extracted from \eref{eq:Intro_NRG}: $\dt$ should be
equal to the displaced charge $\dch$ of \Eq{eq:Intro_Friedel}, which
can also be calculated directly from NRG by calculating the expectation 
value of $\hat{n}_{\tot}$ for $\Gi$ and $\Gf$ individually.\cite{Weichselbaum2011} 
This check was successfully performed, for example, in Refs.~\onlinecite{Helmes2005} 
and \onlinecite{Tureci2011}, within the context of the single impurity
Anderson model; for a recent systematic study, see Ref.~\onlinecite{Weichselbaum2011}. 
We have also performed this check in the present work wherever it was feasible.

Within NRG, it is also possible to directly calculate spectral functions 
such as $\AXw$ of \eref{subeq:SpectralFunction}. To this end, one 
uses two separate NRG runs to calculate the ground state $\Gi$ of 
$\Hi$ and an approximate but complete set of eigenstates $\ket{n}$ 
of $\Hf$. \cite{Anders2005,Anders2006} The Lehmann sum in 
\eref{subeq:SpectralFunction} can then be evaluated explicitly, 
\cite{Peters2006,Weichselbaum2007} while representing the 
$\delta$-functions occurring therein using a log-Gaussian 
broadening scheme. To this end, we follow the approach of 
Ref.~\onlinecite{Weichselbaum2007}, which involves a broadening 
parameter $\sigma$. (The specific choice of NRG parameters $\Lambda$, 
$N_{k}$ and $\sigma$ used for spectral data shown below will be 
specified in the legends of the corresponding figures.) That this 
approach is capable of yielding spectral functions whose asymptotic 
behaviour shows power-law behaviour characteristic of AO has been 
demonstrated recently in the context of the Kondo exciton problem. 
\cite{Helmes2005,Tureci2011,Latta2011} In the examples to be 
discussed below, we will compare the power-law exponents extracted 
from the asymptotic behaviour of such spectral functions to the values 
expected from AO, thus checking relations such as \eref{eq:Intro_A_X} 
for $\AXw$ and \eref{eq:Hopfield_dY} for $\A_{Y} (\omega)$.

%---------------------------------------------------------------------
\section{Interacting Resonant Level Model}\label{sec:IRLM}
%---------------------------------------------------------------------

In this section we consider the effect of AO on dynamical quantities 
in the context of the spinless interacting resonant level model (IRLM). \cite{Gogolin,Schlottmann1980}
(The effects of AO for some static properties of this model were studied 
in Ref.~\onlinecite{Borda2008}.) The purpose of this exercise is to 
illustrate several effects that will be found to arise also for more 
complex models considered in subsequent sections. The IRLM involves 
a single localized level, to be called dot level (alluding to localized 
levels in a quantum dot), interacting with and tunnel-coupled to a 
single Fermi sea. We consider first the case without tunnelling, in 
which case the IRLM reduces to the LCM introduced in \sref{sec:AOconsequences} 
above, where adding an electron to the dot at time $t=0$ constitutes 
a type 2 quench. This leads to AO between the initial and final ground 
states, and corresponding nontrivial AO power laws, $\omega^{-1 + 2\eta}$, 
in spectral functions. We then turn on tunnelling, which connects the 
sectors of Hilbert space for which the dot is empty or filled, and hence 
counteracts AO. Correspondingly, the power-laws get modified at 
frequencies smaller than the renormalized level width, $\omega \lesssim 
\Gren$, where the AO behaviour is replaced by simple Fermi liquid behaviour; 
the effects of AO do survive, however, in a regime of intermediate frequencies, 
$\Gren < \omega < D$.  Finally, we consider quenches of the position of 
the dot level, in which case AO reemerges.

%---------------------------------------------------------------------
\subsection{Without tunnelling: LCM}\label{sec:IRLM_A}
%---------------------------------------------------------------------

\begin{figure}
\begin{centering}
\includegraphics{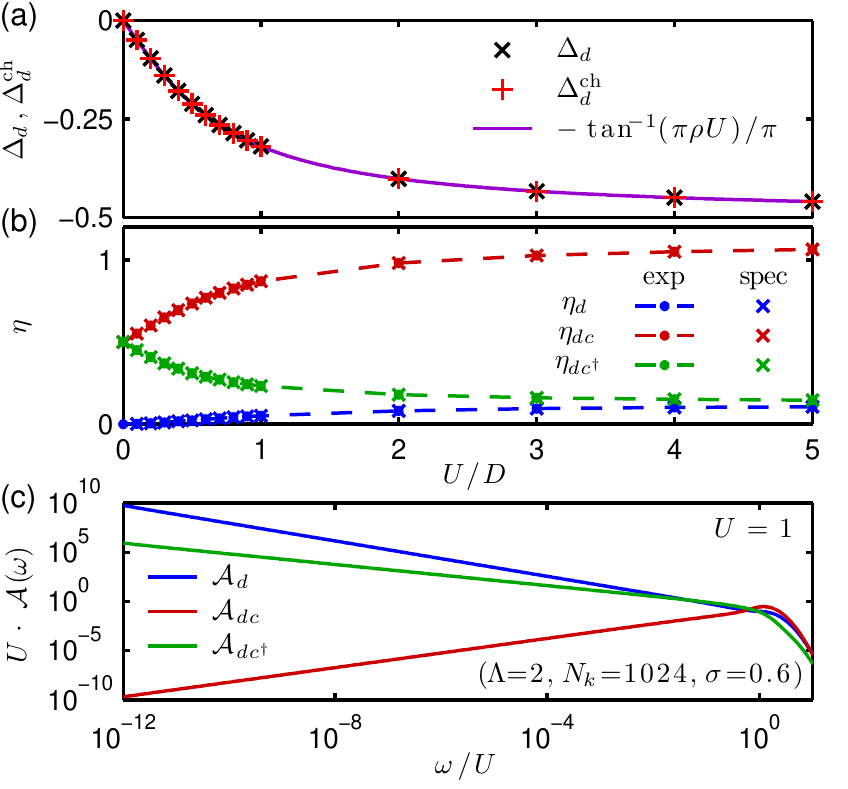}
\par
\end{centering}
\caption{\label{fig:IRLM_A-results} (Color online) 
Numerical results for the LCM of \eref{eq:LCM}, for the type 2 quench of 
\eref{eq:GiGfLCMtype2}, whose initial, final and post-quench initial states 
$\Gi$, $\Gf$ and $\ket{\psi_{\initial}}$ are depicted in Figs.~\ref{fig:IRLM_A}(b,c), 
\ref{fig:IRLM_A}(d) and \ref{fig:IRLM_A}(e-g), respectively. 
(a) Comparison of the decay exponent $\dtd$ obtained from \eref{eq:SIRL_A_Odd} 
(crosses) with the displaced charge $\dtd^{\charge}$ from \eref{eq:displaced-charge-lcm} 
(pluses), for a number of different values of $U$. The two values agree very well 
(they differ by less than $0.1\%$), also with the analytic prediction \eref{eq:delta_analytical} 
(solid line). As expected, $\dtd \to -1/2$ for $U \to \infty$. 
(b) Comparison of two ways of determining the AO exponents $\eta$ that 
govern the low-energy asymptotic behaviour $\A \sim \omega^{-1 + 2\eta}$ 
of the spectral functions of \erefs{eq:SIRL_A_AOspec}, related to Figs.~\ref{fig:IRLM_A}(e-g): 
exponents obtained by fitting a power law to the corresponding spectra [shown in (c)] 
are shown as crosses (marked \qm{spec}, for \qm{spectra}); the corresponding 
exponents expected from \eref{eq:SIRL_A_AOspec}, using the results of (a) for 
$\dtd$, are shown as dots (marked \qm{exp} for \qm{expected}).  We find a 
maximal deviation of less than $1\%$. Here and in all similar figures below, 
the dashed lines are only guides to the eye. 
(c) Asymptotic low-frequency dependence of the spectra \erefs{eq:SIRL_A_AOspec}, 
for $U=1$, on a double logarithmic plot, allowing the corresponding exponents 
$\eta$ to be extracted. 
}
\end{figure}

In this subsection we present numerical results for the IRLM without
tunnelling, corresponding to the local charge model of \eref{eq:LCM},
depicted in \frefsub{fig:IRLM_A}{a}. We consider the type 2 quench of
\eref{eq:GiGfLCMtype2}, with $\hat{X}^{\dagger} = \hat{d}^{\dagger}$.
The initial and final ground states $\Gi$ and $\Gf$ are illustrated in
Figs.~\ref{fig:IRLM_A}(b,c) and \ref{fig:IRLM_A}(d), respectively, and the 
post-quench initial state  $\ket{\psi_{\initial}} = \hat{d}^\dagger \Gi$ 
in \frefsub{fig:IRLM_A}{e}. With these choices the overlap 
$\vert \braket{\psi_{\initial}}{\gfinal} \vert$ of \eref{eq:SIRL_A_Od} becomes
\begin{equation}
	\mathcal{O}_{d} \equiv \vert \bra{G_{0}} \hat{d} \ket{G_{1}} \vert
	= \vert \braket{g_{0}}{g_{1}} \vert \sim N^{-\frac{1}{2} \dtd^2}
	 \label{eq:SIRL_A_Odd} \, .
\end{equation}
The corresponding displaced charge obtained from \eref{eq:Intro_Friedel-X} is 
\begin{equation}
	\dtd^{\charge} = \bra{g_{1}} \hat{n}_{\sea} \ket{g_{1}} - \bra{g_{0}} \hat{n}_{\sea} \ket{g_{0}}
	\komma \label{eq:displaced-charge-lcm}
\end{equation}
since $\Gf$ and $\ket{\psi_{\initial}}$ describe the same dot charge, $n_{d} = 1$. 

We used NRG to calculate the overlap $\mathcal{O}_{d}$ of
\eref{eq:SIRL_A_Odd} and extract the exponent $\dtd$ from its
exponential decay with Wilson chain length [\eref{eq:Intro_NRG}], for
several values of $U$. As consistency check, we also calculated the
displaced charge $\dtd^{\charge}$ [\eref{eq:displaced-charge-lcm}].
As shown in \frefsub{fig:IRLM_A-results}{a}, the results for $\dtd$
(crosses) and $\dtd^{\charge}$ (pluses) agree very well. The displaced
charge $\dtd^{\charge}$ is $< 0$, since the repulsive interaction
pushes charge away from the local site.  Its magnitude $\vert
\dtd^{\charge} \vert$ depends on the interaction strength: as $U$ is
increased from $0$ to $\infty$, the displaced charge goes from $0$ to
$-\frac{1}{2}$, reflecting the complete depletion of the initially
half-filled Wilson chain site directly adjacent to the dot site
[compare Figs.~\ref{fig:IRLM_A}(b) and \ref{fig:IRLM_A}(d)]. 
\Frefsub{fig:IRLM_A-results}{a} shows that the numerical results for
$\dtd$ and $\dtd^{\charge}$ (symbols) also agree with the analytical
result (solid line) obtained for the phase shift obtained from
elementary scattering theory [see e.g.\ Ref.~\onlinecite{Gogolin},
Eq.~(25.29)],
\begin{eqnarray}
	\dtd = - \frac{1}{\pi} \atan \left( \pi \rho U \right)
	\komma \label{eq:delta_analytical}
\end{eqnarray}
with $\rho$ the density of states in the Fermi sea (cf.\ \sref{sec:NRG}). 

To study the influence of AO on dynamical quantities, we consider simple 
and composite type 2 quenches induced by acting on the initial ground 
state $\Gi = \ket{G_{0}}$ with the operators
\begin{equation}
	\hat{X}^{\dagger} = \hat{d}^{\dagger} \komma \qquad
	\hat{Y}_{1}^{\dagger} = \hat{c}^{\dagger} \hat{d}^{\dagger} \komma \qquad
	\hat{Y}_{2}^{\dagger} = \hat{c} \hat{d}^{\dagger}
	\punkt \label{eq:defineXYZ}
\end{equation}
All three operators describe transitions between the $n_{d} = 0$ 
and $1$ sectors. The analysis of \sref{sec:AO-spectral-functions}
applies directly, with the identifications $\Hi = \hat{H}_{0}$ and 
$\Hf = \hat{H}_{1}$, while $\dtC = \pm 1$ for $\hat{Y}_{1}^{\dagger}$ 
or $\hat{Y}_{2}^{\dagger}$, respectively [see \frefsubt{fig:IRLM_A}{e}{g}]. 
In particular, \erefs{eq:Intro_A_X} and \erefb{eq:Ay} imply:
\begin{subequations}
\label{eq:SIRL_A_AOspec}
\begin{align}
	\A_{d} ( \omega ) & \sim \omega^{-1 + 2\dd} & \dd
	& = \tfrac{1}{2} \dtd^2
	\komma \label{eq:SIRL_A_Ad} \\
	\A_{dc} ( \omega ) & \sim \omega^{-1 + 2\ddc} & \ddc
	& = \tfrac{1}{2} (\dtd - 1)^2
	\komma \\
	\A_{dc^{\dagger}} ( \omega ) & \sim \omega^{-1 + 2\ddct} & \ddct
	& = \tfrac{1}{2} (\dtd + 1)^2
	\punkt \label{eq:SIRL_A_Adct}
\end{align}
\end{subequations}
Using NRG, we calculated these three spectra for several values of $U$ 
(cf.\ \frefsub{fig:IRLM_A-results}{c}). In the limit of small $\omega$, the 
spectra show clear power law behaviour, $\omega^{-1 + 2\eta}$. The 
exponents $\eta_{d}$, $\eta_{d c}$, $\eta_{d c^{\dagger}}$ extracted from 
these spectra are shown in \frefsub{fig:IRLM_A-results}{b} (crosses, marked 
\qm{spec}, for \qm{spectra}). They agree well with the values expected 
(dots, marked \qm{exp}, for \qm{expected}) from \erefs{eq:SIRL_A_AOspec}, 
based on the value for $\dtd$ extracted from \eref{eq:SIRL_A_Odd}. Thus, 
all ways of determining $\dtd$ are completely consistent, confirming the 
validity of the above analysis. 

%---------------------------------------------------------------------
\subsection{With tunnelling: IRLM}\label{sec:IRLM_B}
%---------------------------------------------------------------------

\begin{figure}
\begin{centering}
\includegraphics{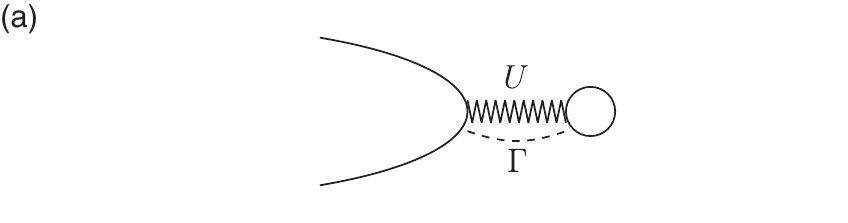}
\includegraphics{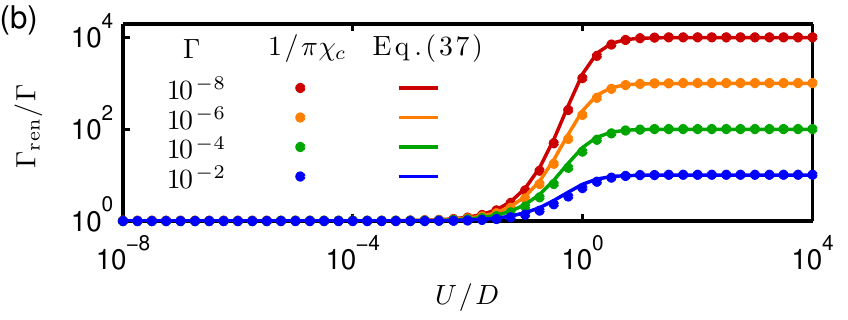}
\par
\end{centering}
\caption{\label{fig:AOC02} (Color online)
(a) Cartoon of the Hamiltonian \erefb{eq:SIRL_B_H} for the IRLM. 
(b) The renormalized level width $\Gren$, calculated via the dot's 
charge susceptibility, $ 1 / \pi\chi_{c}$, \protect\cite{Borda2008} (dots) 
or via \eref{eq:Gren} (solid line), shown as a function of $U$ for 
$\epsd=0$ and several values of $\Gamma$. As $U$ increases 
from 0, $\Gren/\Gamma$ begins to differ significantly from its 
initial value, namely 1, only once $U$ becomes comparable to the 
band-width, reaching its maximal value $(\Gamma/D)^{-1/2}$ 
for $U \gg D$.
}
\end{figure}

The previous subsection focused on a switch between two sectors of the 
Hilbert space, with $n_{d} = 0$ and $n_{d} = 1$, that were not coupled 
dynamically, but governed instead by two distinct Hamiltonians, $\Hi$ 
and $\Hf$. In the present subsection, we consider the case that the sectors 
with $n_{d} = 0$ and $n_{d} = 1$ are coupled by tunnelling between dot 
and lead, so that the notion of an initial and final Hamiltonian, acting in 
decoupled sectors of Hilbert space, does not apply. The dynamics is governed 
instead by the single Hamiltonian $\Hi = \Hf = \hat{H}_{\rm IRLM}$, given 
by [see \frefsub{fig:AOC02}{a}]
\begin{eqnarray}
	\hat{H}_{\rm IRLM} & = & \epsd \, \hat{d}^{\dagger} \! \hat{d}^{\pdag}
	+ U (\hat{d}^{\dagger} \! \hat{d}^{\pdag} -1/2) 
	(\hat{c}^{\dagger} \! \hat{c}^{\pdag} - 1/2) 
	\label{eq:SIRL_B_H} \\
	& & + \sum_{\varepsilon} \varepsilon \, \ced \ce
	+ \sqrt{\frac{\Gamma}{\pi \rho}} \sum_{\varepsilon}
	( \hat{d}^{\dagger} \! \ce + \ced \hat{d}^{\pdag} \!\! )
	\punkt \nonumber
\end{eqnarray}
We assume, here and in all later settings, that the hybridization of the 
dot level with the Fermi sea states is $\varepsilon$-independent, with 
$\Gamma$ being the bare width of the dot level. Here, in contrast to 
the local charge model of \eref{eq:LCM}, the interaction term is taken 
to be particle-hole symmetric, so that the model is particle-hole 
symmetric for $\epsd = 0$. 

The presence of the interaction, $U$, is known to effectively 
modify the level width,\cite{Schlottmann1980,Borda2008} both 
by reducing the density of states of the leads near the dot, and 
by inducing  AO in the leads when the dot occupancy changes. 
The precise interplay between these effects can be quite intricate 
and was studied in Ref.~\onlinecite{Borda2008}. A quantitative 
analysis can be performed by defining a renormalized level width 
in terms of the charge susceptibility, $\Gren \equiv 1 / \pi \chi_{c}$. 
At the point of particle-hole (ph) symmetry ($\epsd = 0$), an analytic 
formula for the latter is available, \cite{Schlottmann1980} 
\begin{eqnarray}
	\Gren / D = ( \Gamma / D )^{ 1 / ( 2 - ( 1 + \dtdph)^2 ) }
	\komma \label{eq:Gren}
\end{eqnarray}
where $\dtdph$ is given by
\begin{eqnarray}
	\dtdph & = & - \frac{2}{\pi} \atan \left( \pi \rho U / 2 \right) 
	\punkt \label{eq:delta_analytical_ph}
\end{eqnarray}
$\dtdph$ can be interpreted as the change in scattering phase shift 
that a system with $\Gamma = 0$, $\epsd = 0$ experiences if the 
local occupancy is changed abruptly from $n_{d} = 0$ to $1$. The 
form of \Eq{eq:delta_analytical_ph} is analogous to \eref{eq:delta_analytical} 
for $\dtd$, with two differences: since the final scattering potentials 
being compared have amplitude $-U/2$ and $U/2$ (instead of $0$ 
and $U$), the argument of $\tan^{-1}$ has an extra factor of $1/2$, 
and there is an extra prefactor of $2$. 

The dependence of $\Gren$ on $U$ is illustrated in \frefsub{fig:AOC02}{b}, 
which shows good agreement between the NRG results for $1 / \pi \chi_{c}$ 
(dots) and the analytic formula \erefb{eq:Gren} (lines). For $U$ much smaller 
than the bandwidth $D$, $\Gren/\Gamma$ is essentially equal to $1$; it 
strongly increases once $U$ becomes of the order $D$, and saturates to 
$(\Gamma/D)^{-1/2}$ for $U \gg D$. 

\begin{figure}
\begin{centering}
\includegraphics{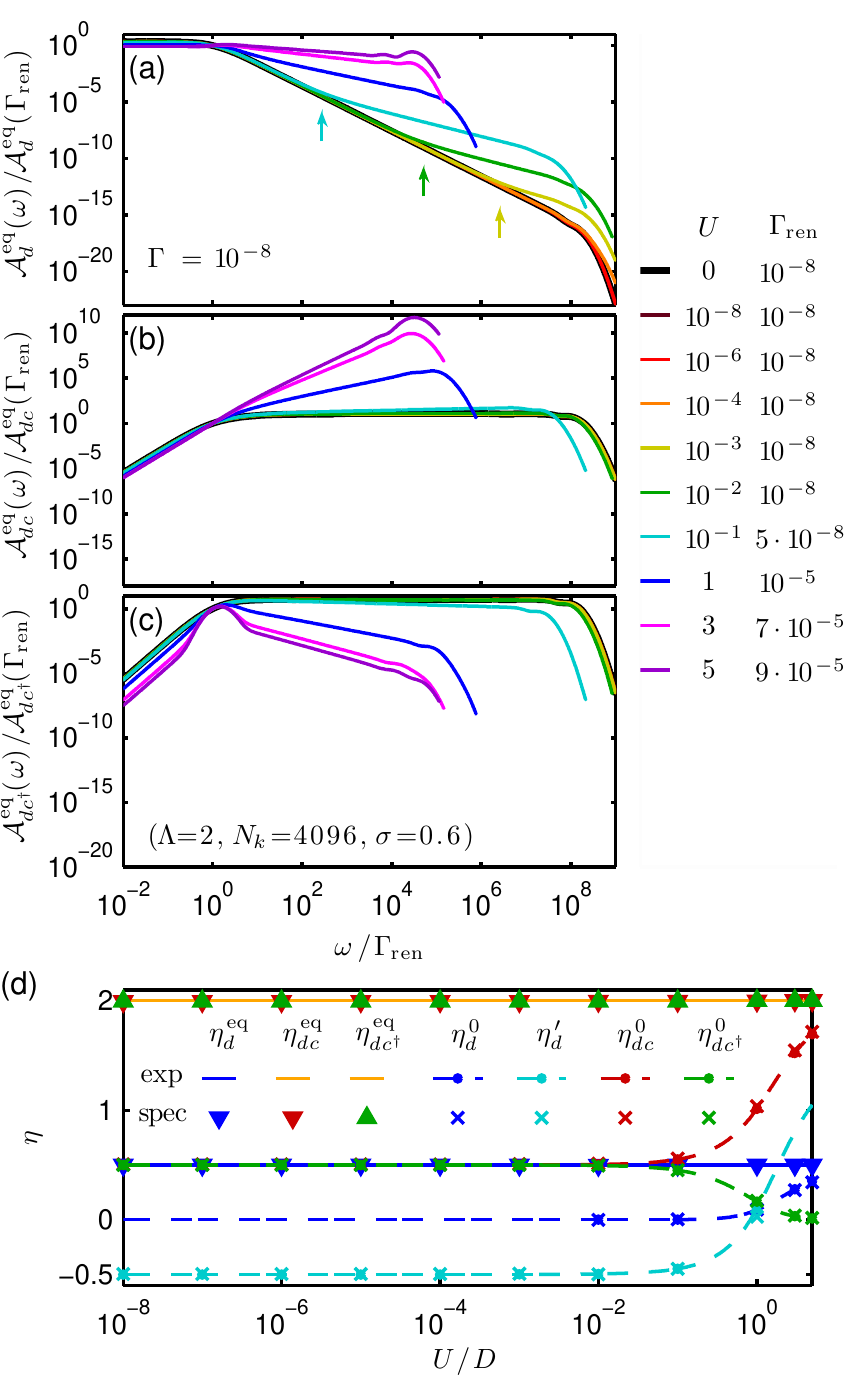}
\par
\end{centering}
\caption{\label{fig:AOC02spec} (Color online)
(a)-(c) The equilibrium spectral functions $\A^{\equil}_{d} (\omega)$, 
$\A^{\equil}_{dc} (\omega)$ and $\A^{\equil}_{dc^{\dagger}} (\omega)$ 
for the IRLM, showing a crossover from trivial power laws, $\omega^{-1 
+ 2 \eta^{\equil}}$, for $\omega < \ostar$, to AO power laws, $\omega^{-1 
+ 2 \etap}$, for $\ostar < \omega < D$, with the crossover frequency 
$\ostar$ given by $\Gren$. 
(d) Comparison of the exponents $\eta^{\equil}$ (triangles) and $\etap$ 
or $\eta_{d}'$ (crosses) extracted from the spectra shown in (a-c), with the 
values expected from \erefs{eq:SIRL_B_FL} for $\eta^\equil$ (solid lines), 
and from \erefs{eq:SIRL_B_AOspec} for $\etap$ or from \eref{eq:Aeqd-Lorentziandecay} 
for $\eta_{d}'$ (dashed lines with dots), for several values of $U$. In (a), 
arrows indicate the scale $\barostar$ that separates the regimes $\ostar 
< \omega < \barostar$ and $\barostar < \omega < D$, where $\A_{d}^\equil$ 
scales according to \Eqs{eq:Aeqd-Lorentziandecay} or \erefb{eq:SIRL_B_AOspec-Ad}, 
respectively. 
}
\end{figure}

Let us now consider the equilibrium spectral functions for the 
operators of \eref{eq:defineXYZ}, $\A^\equil_{d}$, $\A^\equil_{dc}$ 
and $\A^\equil_{dc^{\dagger}}$. They are defined as in \eref{subeq:SpectralFunction} 
but with $\Hf = \Hi$, because for the IRLM, where $n_{d}$ is not conserved, 
none of these operators induces a quench. Therefore, the $\omega \to 0$ 
behaviour of their correlators is expected (and indeed found) to be independent 
of AO. However, quite remarkably, traces of AO \emph{do show up} in an 
intermediate frequency regime, $\ostar < \omega < D$, where $1/\ostar$ 
corresponds to the time scale within which charge equilibration takes place. 
Below the energy scale $\ostar$ the quantum impurity becomes strongly 
correlated with the Fermi sea and for the present model we have $\ostar 
\simeq \Gren$. Let us therefore discuss the two regimes, $\omega$ below 
or above $\ostar$, separately.

In the regime $\omega \ll \ostar$, the spectral functions are found to 
have the following asymptotic form $\A^{\equil} \sim \omega^{-1 + 2 
\eta^{\equil}}$ [cf.\ \frefsubt{fig:AOC02spec}{a}{c}]: 
\begin{subequations}
\begin{align}
	\A^\equil_{d} ( \omega )  & \sim \omega^0
	\komma
	& \eta^\equil_d  & =  1/2
	\komma  \label{eq:SIRL_B_FL-ddagger} \\
 	\A^\equil_{dc} ( \omega ) & \sim \omega^3
	\komma
	& \eta^\equil_{dc} & =  2
	\komma \label{eq:SIRL_B_FL-cdaggerddagger} \\
	\A^\equil_{dc^{\dagger}} ( \omega ) & \sim \omega^3
	\komma
	& \eta^\equil_{dc^\dagger} & = 2
	\punkt \label{eq:SIRL_B_FL-cddagger}
\end{align}
\label{eq:SIRL_B_FL}
\end{subequations}
The exponents arising here can be understood analytically using 
elementary, though not entirely trivial arguments, based on the 
fact that the lowest-lying excitations of this model have Fermi 
liquid properties. We refer the reader to the Appendix for a detailed 
analysis. 

Now consider the regime $\ostar < \omega < D$. As shown in the 
corresponding regime of $\omega/\Gren > 1$ in \frefsubt{fig:AOC02spec}{a}{c}, 
each of the equilibrium spectral functions $\A^\equil_{d}$, $\A^\equil_{d c}$ 
and $\A^\equil_{d c^{\dagger}}$, exhibits another, different power-law there.
For $\A^\equil_{d}$ we actually find that within this regime two different 
power-laws can be discerned: First, in a regime $\ostar < \omega < \barostar$ 
we find, 
\begin{eqnarray}
	\A_{d}^{\equil} (\omega) & \sim \omega^{-1 + 2 \eta_{d}'}
	\; , \quad
	\eta_{d}' = - \tfrac{1}{2} - \dtdph + (\dtdph)^2
	\; , \qqph
	\label{eq:Aeqd-Lorentziandecay}
\end{eqnarray}
where $\dtdph$ is given by \eref{eq:delta_analytical_ph}. The exponent
$\eta_{d}'$ corresponds to the leading correction for weak
interactions $(U/D \ll 1)$ to the pure Lorentzian decay of the
spectral function of the $d$-level (see \eref{eq:Lorenztian-spectral-function} 
of the Appendix), as can be
shown using methods discussed in Refs.~\onlinecite{Goldstein2009} and
\onlinecite{Goldstein2010b}. The scale $\barostar$ that sets the upper
limit for this behaviour is marked by arrows in
\frefsub{fig:AOC02spec}{a} and decreases with increasing $U/D$.  For
$U/D$ sufficiently small that $\barostar$ lies far below the bandwidth
$D$, we find a second power law within the window $\barostar < \omega
< D$, namely
\begin{subequations}
\label{eq:SIRL_B_AOspec}
\begin{align}
	\label{eq:SIRL_B_AOspec-Ad}
	\A^\equil_{d} ( \omega ) & \sim \omega^{-1 + 2\ddp} ,
	& \ddp & = \tfrac{1}{2} (\dtdph)^2
	% \quad (U \gtrsim D)
	\punkt
\end{align}
For the other two spectral functions we find throughout the regime 
$\ostar < \omega < D$: 
\begin{align}
	\A^\equil_{dc} ( \omega )  & \sim \omega^{-1 + 2\ddcp} ,
	& \ddcp & = \tfrac{1}{2} (\dtdph - 1)^2
	\komma \\
	\A^\equil_{dc^{\dagger}} ( \omega ) & \sim \omega^{-1 + 2\ddctp} ,
	& \ddctp & = \tfrac{1}{2} (\dtdph + 1)^2
	\punkt \label{eq:SIRL_B_AOspec_dcdagger}
\end{align}
\end{subequations}
Remarkably, \erefs{eq:SIRL_B_AOspec} have the same form as 
\erefs{eq:SIRL_A_AOspec}, except that $\dtd$ is replaced by 
$\dtdph$ of \eref{eq:delta_analytical_ph}, i.e.\ by the AO 
exponent involved in abruptly changing the local occupancy 
from $0$ to $1$ (in the absence of tunnelling). That this 
exponent should emerge is natural, since the corresponding 
correlators $\G_{d}$, $\G_{dc}$ and $\G_{dc^\dagger}$ all 
involve an operator $d^\dagger$ that places an electron on 
the dot at time $t=0$. Although the dot occupancy $n_d (t)$ 
will relax back to its initial value $n_{d}^{\initial}$ in the long 
time limit, this requires times $t \gg 1/\ostar$. In contrast, 
the lead electrons react to the change in local charge on the 
much shorter time scale $1/D$. Thus, in the window of intermediate 
times, $1/D \ll t \ll 1/\ostar$, corresponding to frequencies 
$\ostar \ll \omega \ll D$, the situation is similar to that of 
the previous subsection, where we had $\Gamma = 0$ and 
a change in dot occupancy from $0$ to $1$ induced changes 
in the lead phase shifts, accompanied by AO. Thus, the exponents 
$\etap$ arising in \eref{eq:SIRL_B_AOspec} can be identified as the 
(equilibrium) scaling dimensions of the corresponding operators 
calculated in the \emph{absence} of tunnelling (which is why 
we use a superscript $0$ on such exponents, here and below). 
This explains the similarity between the behaviour described by 
\erefs{eq:SIRL_B_AOspec} and \erefs{eq:SIRL_A_AOspec}. Note that 
the scaling dimension $\ddctp$   [\eref{eq:SIRL_B_AOspec_dcdagger}] 
of the tunnelling operators $\hat{d} \hat{c}^{\dagger}$ and $\hat{c}^\dagger 
\! \hat{d}$ satisfy $0 \le \ddctp \le 1/2$ [since for $U > 0$, we have 
$-1 \le \dtdph \le 0$, by \eref{eq:delta_analytical_ph}], thus tunnelling 
is always relevant for this model. 

We conclude this subsection with a comment on the fact that $\A^\equil_{d} 
(\omega)$ crosses over from non-AO behaviour [\eref{eq:Aeqd-Lorentziandecay}] 
to AO behaviour [\eref{eq:SIRL_B_AOspec-Ad}] as $U/D$ is increased past 1. 
AO behaviour is absent for $U/D \ll 1$ because this situation corresponds 
essentially to a non-interacting resonant-level model, for which $\G_{d}^\equil 
(t)$ does \emph{not} show power-law behaviour of the type assumed in 
\eref{eq:Intro_Correlator2}; instead it decays exponentially ($\sim e^{-\Gamma t}$), 
causing the spectral function $\A^\equil_{d} (\omega)$ to have an essentially 
Lorentzian form. \Eref{eq:Aeqd-Lorentziandecay} is the large-frequency limit 
of the latter, but including the leading corrections in $U/D$, calculated using 
methods discussed in Refs.~\onlinecite{Goldstein2009} and~\onlinecite{Goldstein2010b}. 
However, once $U/D$ becomes $\gtrsim 1$, AO does begin to matter, implying 
a regime of power-law decay for $\G_{d}^\equil (t)$ on intermediate time scales, 
leading to \eref{eq:SIRL_B_AOspec-Ad} for $\A^{\equil}_{d} (\omega)$. 

%---------------------------------------------------------------------
\subsection{Quantum quench of level position}\label{sec:IRLM_C}
%---------------------------------------------------------------------

\begin{figure}
\begin{centering}
\includegraphics{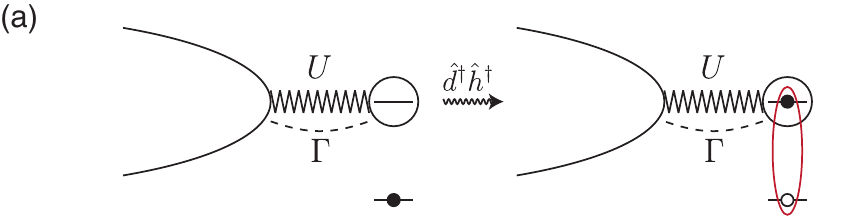}
\includegraphics{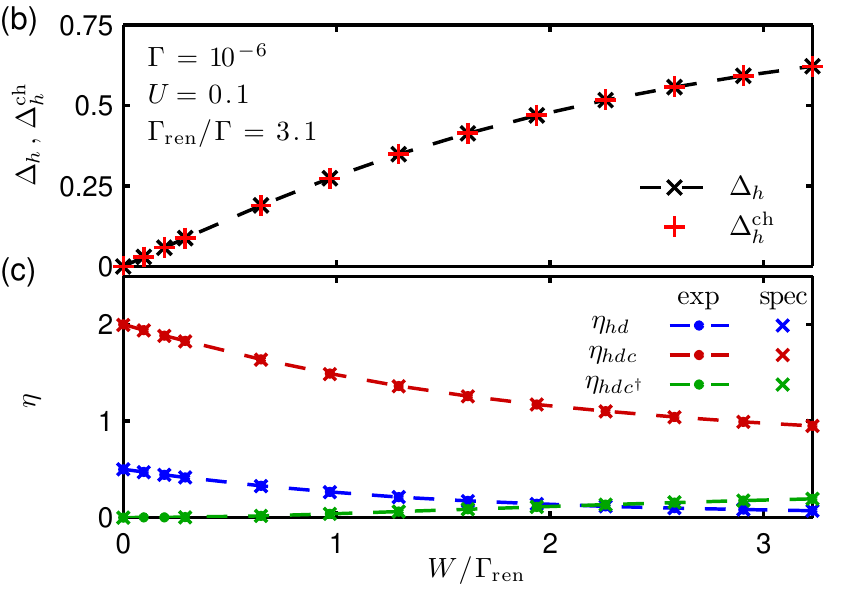}
\par
\end{centering}
\caption{\label{fig:AOC03} (Color online) 
(a) Cartoon of the quench which occurs when an electron-hole pair 
is created at time $t=0$, see \eref{eq:IRLM_C_H}. (The cartoon depicts 
the situation relevant  for exciton creation by absorption of a photon, 
which excited an electron from a valence-band to a conduction band 
level of a semiconducting quantum dot.) 
(b) The exponent $\dth$ [from \eref{eq:SIRL_A_Od}] and the displaced 
charge $\dth^{\charge}$ [from \eref{eq:Intro_Friedel-X}], for the quench 
of \eref{eq:quench-by-W}, as function of the quench range $W$. 
(c) Corresponding values of the AO exponents $\eta_{h d}$, $\eta_{h d c}$ 
and $\eta_{h d c^{\dagger}}$, extracted from the asymptotic behaviour 
$\omega^{-1 + 2\eta}$ of spectral functions (crosses), or as expected 
from \erefs{eq:IRLM_C_A-d-dc-dct} (dots). Typically, relative errors are 
less than $1\%$. 
}
\end{figure}

\begin{figure}
\begin{centering}
\includegraphics{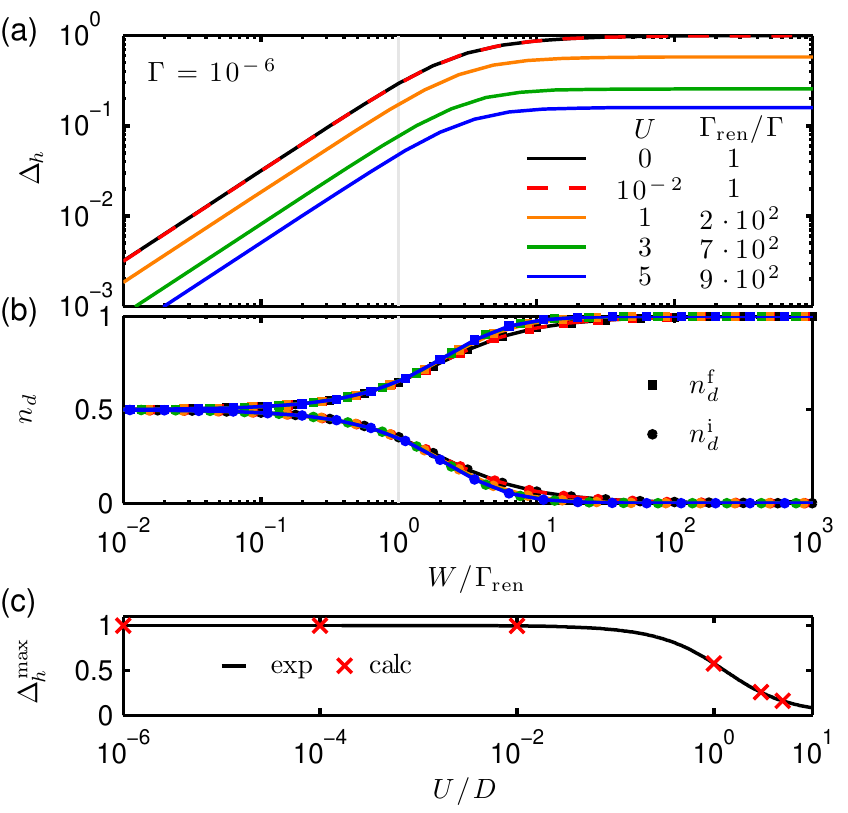}
\par
\end{centering}
\caption{\label{fig:AOC03_Sweep} (Color online) 
(a) The AO exponent $\dth$ [extracted according to \eref{eq:SIRL_A_Od}] 
as function of $W/\Gren$, for several values of $U$. For $W / \Gren \gg 1$, 
$\dth$ approaches its maximal value $\dth^{\rm max}$. As $W$ is 
reduced below $\Gren$, $\dth$ drops below its maximal value and 
decreases with $W$ [linearly so for $W / \Gren \ll 1$]. 
(b) The final and initial occupancies $n_d^\final$ (squares, upper curves) 
and $n_d^\initial$ (circles, lower curves) as functions of $W/\Gren$, for 
the same values of $U$ [same color code as in (a)].
(c) The maximal value $\dth^{\rm max}$  of the AO exponent $\dth$, 
extracted from the $W / \Gren \gg 1$ regime of (a) (crosses, \qm{calc}), or 
expected from \eref{eq:dth_max} (solid line, \qm{exp}); the relative deviations 
are well below $1\%$. 
}
\end{figure}

In the previous subsection we emphasized the importance of the scale 
$\ostar$, which separates the low- and intermediate-frequency regimes, 
showing trivial exponents or AO exponents, respectively. It is instructive 
to study the role of the scale $\ostar$ in a slightly different but related 
context, namely quench spectral functions involving a quantum quench 
of the level position. This will shed further light on the AO between states 
with different local level occupancies. 

Concretely, we consider initial and final Hamiltonians that both 
are of the form \eref{eq:SIRL_B_H}, but with initial and final level 
positions that are symmetrically spaced on opposite sides of the 
model's symmetry point at $\epsd = 0$:
\begin{eqnarray}
	\epsd^{\initial} = W/2
	\quad \stackrel{{\rm quench}}{\longrightarrow} \quad
	\epsd^{\final}  =  - W/2
	\punkt \label{eq:quench-by-W}
\end{eqnarray}
Although this is an example of a type 1 quench, it will be convenient 
(mainly for notational reasons) to reformulate this situation as a type 
2 quench. To this end we use the Hamiltonian 
\begin{eqnarray}
	\hat{H} & = &
	W (1/2 -  \hat{n}_{h} ) \hat{n}_{d}
	+ U (\hat{n}_{d}  -1/2) (\hat{c}^{\dagger} \! \hat{c} - 1/2)
	\nonumber \qqph \\
	& & + \sum_{\varepsilon} \varepsilon \, \ced \ce
	+ \sqrt{\frac{\Gamma}{\pi\rho}} \sum_{\varepsilon} ( \hat{d}^{\dagger} \! \ce + \ced \hat{d} )
	\komma \label{eq:IRLM_C_H}
\end{eqnarray}
where we have introduced an auxiliary degree of freedom, called 
\qm{hole} (in analogy to the role of holes in exciton creation by 
optical absorption\cite{Helmes2005,Tureci2011,Latta2011}), with 
hole counting operator $\hat{n}_{h} = \hat{h}^{\dagger} \hat{h}$. 
The hole has no dynamics; its only role is to distinguish two distinct 
sectors of Hilbert space, in which the dynamics is described by $\Hi$ 
or $\Hf$, with hole number $n_{h} = 0$ or $1$, respectively [see 
\frefsub{fig:AOC03}{a}].  The type 2 quench that switches between 
these sectors is induced by $\hat{X}^{\dagger} = \hat{h}^{\dagger}$. 
The overlap $\mathcal{O}_{h} \sim N^{-\frac{1}{2} \dth^2}$ between 
the initial and final ground states is characterized by an AO exponent 
$\dth$ [\eref{eq:SIRL_A_Od}] that is equal to the charge $\dth^{\charge}$ 
displaced by the quench [\eref{eq:Intro_Friedel-X}]. 

The magnitude of $\dth$ increases with the range $W$ of the 
quench, as shown in \frefsub{fig:AOC03}{b} (linear scale) and 
\frefsub{fig:AOC03_Sweep}{a} (log-log scale). Note, in particular, 
that the scale on which the quenching range, $W$, needs to change 
in order for the AO exponents to change significantly, is given by 
$\Gren$. This is natural: when $W \gg \Gren$, the two states 
$\Gi$ and $\Gf$ connected by the quench describe dots with 
strongly different occupancies, $n_{d}^{\initial} \simeq 0$ vs.\ 
$n_{d}^{\final} \simeq 1$, see \frefsub{fig:AOC03_Sweep}{b}. 
Hence the AO [\eref{eq:SIRL_A_Od}] of the corresponding Fermi 
seas will be strong. The maximum possible value of the exponent 
$\dth$ is 
\begin{eqnarray}
	\dth^{\rm max} = 1 + \dtdph
	\komma \label{eq:dth_max}
\end{eqnarray}
with $\dtdph ( U ) < 0$ given by \eref{eq:delta_analytical_ph}. The 
first term simply gives the $U \to \infty$ value of the change in dot 
occupancy induced by the quench, namely $1$; the second term 
reflects the reaction of the Fermi sea to this change, cf.\ \sref{sec:IRLM_B}.

Following the arguments of \sref{sec:AO-spectral-functions}, the 
nonequilibrium spectral functions $\A_{Y} (\omega)$, defined for 
\begin{equation}
	\hat{Y}_{1}^{\dagger} = \hat{d}^{\dagger} \hat{h}^{\dagger} \komma \qquad
	\hat{Y}_{2}^{\dagger} = \hat{c}^{\dagger} \hat{d}^{\dagger} \hat{h}^{\dagger} \komma \qquad
	\hat{Y}_{3}^{\dagger} = \hat{c} \, \hat{d}^{\dagger} \hat{h}^{\dagger}
	\komma \label{eq:IRLM_C_defineY}
\end{equation}
are expected to show the following AO behaviour for ${\omega \to 0}$:
\begin{subequations}
\begin{align}
	\A_{h d} ( \omega ) & \sim \omega^{-1 + 2\dhd} & \dhd & = \tfrac{1}{2} ( \dth - 1 )^{2}
	\komma \\
	\A_{h d c} ( \omega ) & \sim \omega^{-1 + 2\dhdc} & \dhdc & = \tfrac{1}{2} ( \dth - 2 )^{2}
	\komma \\ 
	\A_{h d c^{\dagger}} ( \omega ) & \sim \omega^{-1 + 2\dhdct} & \dhdct & = \tfrac{1}{2} \dth^2
	\punkt
\end{align}
\label{eq:IRLM_C_A-d-dc-dct}
\end{subequations}
The reason for the specific form of the exponents is that for the 
correlators $\G_{h d}$, $G_{h d c}$ or $G_{h d c^{\dagger}}$, at 
$t=0$ the local charge (on the d-level or in the Fermi sea) is 
increased by one, two or zero, respectively [i.e.\ $\dtC = 1$, $2$ 
or $0$ in \eref{eq:related-displaced-charges}]. \Frefsub{fig:AOC03}{c} 
shows that the exponents (crosses) extracted from the asymptotic 
behaviour $\A_Y (\omega)$ are indeed in good agreement with 
values expected (dots) from \erefs{eq:IRLM_C_A-d-dc-dct}. 

%---------------------------------------------------------------------
\section{Population Switching without Sensor}\label{sec:PS}\label{sec:PS_A}
%---------------------------------------------------------------------

The models investigated so far served as testing ground for the
influence of AO on various types of spectral functions. The following
two sections have the concrete motivation to clarify the role of AO in
the context of quantum dot models that display the phenomenon of
population switching (PS). \cite{Hackenbroich1997,Baltin1999,
  Silvestrov2000,Silvestrov2001,Golosov2006,Karrasch2007,Goldstein2010,Goldstein2011}
In such models, a quantum dot, tunnel-coupled to leads, contains
levels of different widths and is capacitively coupled to a gate
voltage that shifts the levels energy relative to the Fermi level of
the leads. Under suitable conditions, an (adiabatic) sweep of the gate
voltage induces an inversion in the population of these levels (a
so-called population switch), implying a change in the local potential
seen by the Fermi seas in the leads. Goldstein, Berkovits and Gefen
(GBG) have argued in Ref.~\onlinecite{Goldstein2010,Goldstein2011} that in this
context AO can play an important role.  In particular, they pointed
out that for a model involving a third lead acting as a charge sensor,
the effects of AO can be enhanced to such an extent that population
switching becomes abrupt, i.e.\ turns into a phase transition. Our
goal is to elucidate the influence of AO by using the tools developed
above in the context of the IRLM.

In the present section, we will study population switching in a 
two-lead model (without charge sensor), which is equivalent to 
an anisotropic Kondo model. \cite{Goldstein2010,Goldstein2011,
Kashcheyevs2007,Lee2007,Silvestrov2007,Kashcheyevs2009} 
The corresponding Kondo temperature, 
$\TK$, sets the width of the population switch as function of gate 
voltage. We calculate the spectral function $\A_Y^\equil (\omega)$ 
of the pseudospin-flip operator and show that $\TK$ also acts as 
the crossover scale $\ostar$ that separates a low-frequency regime 
showing Fermi-liquid power laws from an intermediate-frequency 
regime revealing AO exponents. We investigate the origin of the 
latter by a quantum quench analysis similar to that of \sref{sec:IRLM_C} 
above. In the following section, we will generalize the model by adding 
a charge sensor and analyse how this enhances the effects of AO. 

%---------------------------------------------------------------------
\subsection{Width of switching regime}\label{sec:PSwidth}
%---------------------------------------------------------------------

\begin{figure}
\begin{centering}
\includegraphics{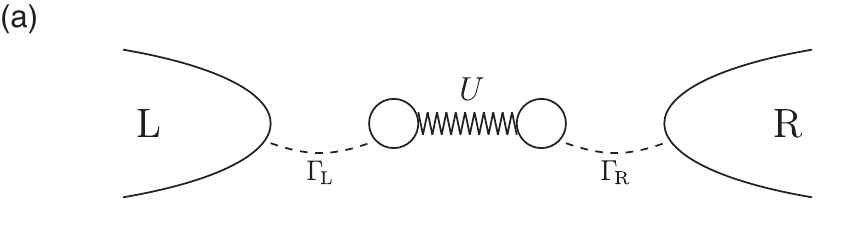}
\includegraphics{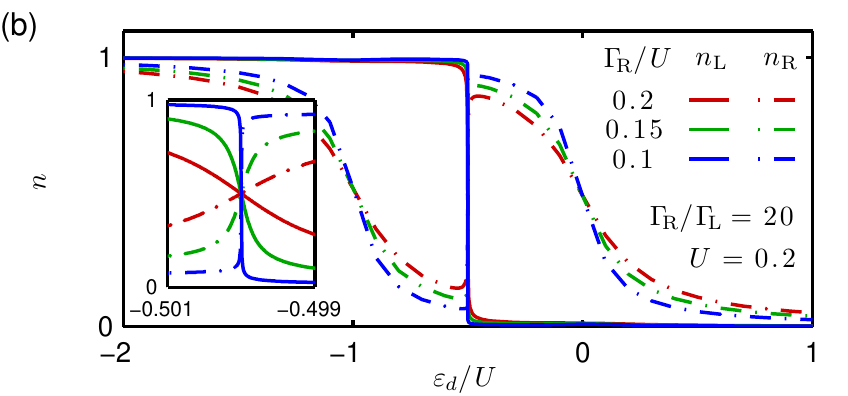}
\includegraphics{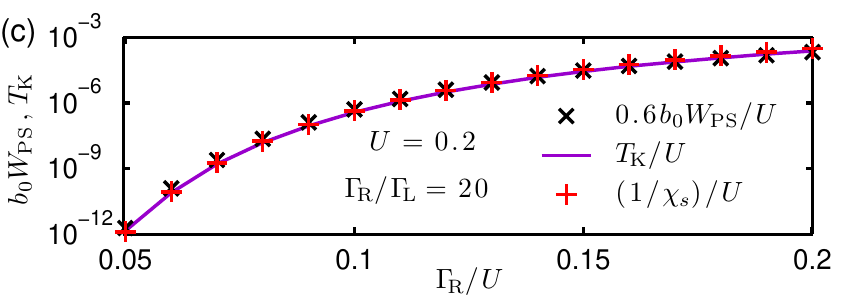}
\par
\end{centering}
\caption{\label{fig:PS_A} (Color online) 
(a) Cartoon of the Hamiltonian \erefb{eq:SIAM} for the asymmetric SIAM. 
(b) The occupations $n_{\rm L}$ (solid lines) and $n_{\rm R}$ (dashed lines) 
of the left and right level, respectively, as functions of $\epsd$, for several 
values of $\GammaR$, at a fixed ratio of $\GammaR / \GammaL = 20$. As 
$\epsd$ is lowered past the particle-hole  symmetric point at $\epsd = -U/2$, 
population switching occurs,  with $n_R$ changing from near $1$ to near $0$, 
and vice versa for $n_{\rm L}$. Inset: zoom into the switching region around 
$\epsd = -U/2$, showing that the switch is continuous (as function of $\epsd$) 
even though the switching region becomes narrower for decreasing $\GammaR$. 
(c) Comparison of $b_{0} \DPS$ [from \erefs{eq:DPS} and \erefb{eq:KondoB0}, 
crosses], $\TK$ [from \eref{eq:TK}, solid], and the inverse pseudospin susceptibility 
$1/ \chi_{s}$ (pluses). All three quantities evidently decrease similarly with decreasing 
$\Gamma_R/U$. 
} 
\end{figure}

We consider a model involving two single-level dots (${\mu = {\rm L,R}}$) 
and for convenience choose their level energies $\epsd$ to be equal, so that 
the PS always occurs at the particle-hole symmetric point, $\epsd = -U/2$. 
(Note that PS occurs also for nondegenerate levels, as long as their level spacing 
is smaller than the difference of their level widths $\Gammamu$.) The levels 
have an electrostatic coupling $U>0$ and are each tunnel-coupled to its own 
lead [see \frefsub{fig:PS_A}{a}]:
\begin{eqnarray}
	\hat{H}_{\rm SIAM} & = &\sum_{\mu}
	\epsd \, \hat{d}_{\mu}^{\dagger} \hat{d}_{\mu}^{\pdag}
	+ U \hat{d}_{\rm L}^{\dagger} \hat{d}_{\rm L}^{\pdag}
	\hat{d}_{\rm R}^{\dagger} \hat{d}_{\rm R}^{\pdag}
	\label{eq:SIAM} \\
	& + & \sum_{\varepsilon \mu} \varepsilon \, \cemd \cem
	+ \sum_{\mu} \sqrt{\frac{\Gammamu}{\pi\rho}}
	\sum_{\varepsilon} ( \hat{d}_{\mu}^{\dagger} \cem + \cemd \hat{d}_{\mu}^{\pdag} )
	\komma \nonumber
\end{eqnarray}
(We use notation analogous to that of \sref{sec:IRLM_B}.) We choose 
the level widths to be strongly asymmetric and will use a fixed value 
of their ratio, $\GammaR / \GammaL = 20$, throughout. The model 
thus has the form of a spin-asymmetric single-impurity Anderson 
model (SIAM), where $\mu$ acts as pseudospin index. 

As illustrated in \frefsub{fig:PS_A}{b}, this model shows PS when $\epsd$ 
is decreased past $\epsd = -U/2$ (the particle-hole symmetric point): 
as this \qm{switching point} is crossed, the occupancy of the broad level 
(dashed lines) changes from near $1$ to near $0$, and vice versa for the 
narrow level (solid lines). We define the width of the switching regime, 
$\DPS$, as the difference, 
\begin{equation}
	\DPS \equiv \epsd (n_{\rm R+})  - \epsd (n_{\rm R-}) 
	\komma \label{eq:DPS}
\end{equation}
between those two values of $\epsd$, located symmetrically on either 
side of the switching point, at which the occupation of the right level 
is $n_{\rm R+} \equiv \tfrac{3}{4} n_{\rm R+}^{\rm max}$ $(> \tfrac{1}{2})$ or 
$n_{\rm R-} = 1 - n_{\rm R+}$ ($ < \tfrac{1}{2})$, respectively, where 
$n_{\rm R+}^{\rm max}$ is the largest value reached by $n_{\rm R}$ for 
$\epsd > -U/2$, to the right of the PS.

\Frefsub{fig:PS_A}{b} and its inset show that the width of the
switching regime decreases with decreasing $\Gammamu$, without,
however, dropping to zero as long as the level widths are nonzero. 
This behaviour can be understood as follows. \cite{Goldstein2010,
Goldstein2011,Kashcheyevs2007,Lee2007,Silvestrov2007,Kashcheyevs2009} 
In the vicinity of the particle-hole symmetric point, only two local 
charge configurations are relevant, namely those with occupancies 
$(n_{\rm L},n_{\rm R})$ equal to $(0,1)$ or $(1,0)$. The spin-asymmetric 
SIAM can thus be mapped onto an anisotropic Kondo model by a 
Schrieffer-Wolff transformation. This leads to an anisotropic pseudospin 
exchange interaction of the form 
\begin{eqnarray}
	\hat{H}_{\rm exch} & =  &
	( U_{\rm L} + U_{\rm R} )
	( \hat{d}_{\rm L}^{\dagger} \hat{d}_{\rm L}^{\pdag} - \hat{d}_{\rm R}^{\dagger} \hat{d}_{\rm R}^{\pdag} )
	( \hat{c}_{\rm L}^{\dagger} \hat{c}_{\rm L}^{\pdag} - \hat{c}_{\rm R}^{\dagger} \hat{c}_{\rm R}^{\pdag} )
	\nonumber \\
	& & + 2 \sqrt{ U_{\rm L} U_{\rm R} }
	( \hat{c}_{\rm L}^{\pdag} \hat{c}_{\rm R}^{\dagger} \hat{d}_{\rm L}^{\dagger} \hat{d}_{\rm R}^{\pdag} + {\rm h.c.})
	\nonumber \\
	& & + \Beff ( \hat{d}_{\rm L}^{\dagger} \hat{d}_{\rm L}^{\pdag} -
	\hat{d}_{\rm R}^{\dagger} \hat{d}_{\rm R}^{\pdag} )/2
	\label{eq:KondoHamilton} \komma
\end{eqnarray}
respectively, with coupling constants given by 
\begin{equation}
	\rho U_{\!\mu} = \frac{\Gammamu}{\pi}
	\left( \frac{1}{\epsd + U} + \frac{1}{\vert \epsd \vert} \right)
	\komma \label{eq:U_alpha}
\end{equation}
and effective magnetic field
\begin{equation}
	\Beff = b_{0} ( \epsd + U/2 ), \quad b_{0} = \frac{4 ( \GammaR - \GammaL ) }{ \pi U }
	\punkt \label{eq:KondoB0}
\end{equation}
The corresponding Kondo temperature is given by the following
expression: \cite{Kashcheyevs2007}
\begin{equation}
	\TK = \frac{\sqrt{U ( \GammaL + \GammaR ) }}{\pi}
	\exp \left[ \frac{\pi \epsd (U + \epsd)}{2U (\GammaL - \GammaR)} \ln \frac{\GammaL}{\GammaR} \right]
	\punkt \label{eq:TK}
\end{equation}
Note that $\TK$ decreases exponentially if $\Gammamu$ is decreased 
with a fixed ratio of $\GammaR / \GammaL$ and actually becomes zero 
for $\Gammamu = 0$ (the argument of the exponent in \Eq{eq:TK} is 
negative, since $\epsd <0$). 

Now, $\TK$ can be associated with the energy gained by forming 
a ground state involving a screened local pseudospin, which in the 
present setting translates to a ground state involving a coherent 
superposition of configurations with local occupancies $(0,1)$ and 
$(1,0)$. Screening will cease when $\epsd$ deviates sufficiently 
from the symmetry point $-U/2$ that the effective magnetic field 
$|\Beff|$ exceeds $\TK$, in which case the ground state will be 
dominated solely by the $(0,1)$ or $(1,0)$ configuration, instead 
of involving a coherent superposition of both. Thus the switching 
width will be set by $b_{0} \DPS \simeq \TK$, up to a numerical 
constant of order unity. 

\Frefsub{fig:PS_A}{c} confirms this expectation. It shows that $b_0$ 
times the switching width $\DPS$ [from \Eq{eq:DPS}] (crosses) and 
the Kondo temperature $\TK$ at $\epsd = -U/2$ [from \Eq{eq:TK}] 
(solid line), when plotted as functions of $\GammaR / U$ at fixed 
$\GammaR / \GammaL$, are indeed almost perfectly proportional 
to each other. As a numerical consistency check, \frefsub{fig:PS_A}{c} 
also shows the inverse of the zero-temperature pseudospin susceptibility 
of the dot levels, $1 / \chi_{s}$ (pluses), confirming that $\TK = 1/ \chi_{s}$. 
(This is analogous to the relation $\Gren = 1 / \pi \chi_{c}$ of \sref{sec:IRLM_B}.) 

%---------------------------------------------------------------------
\subsection{AO in dynamics of pseudospin-flip operator} \label{sec:AO-AY-withoutsensor}
%---------------------------------------------------------------------

\begin{figure}
\begin{centering}
\includegraphics{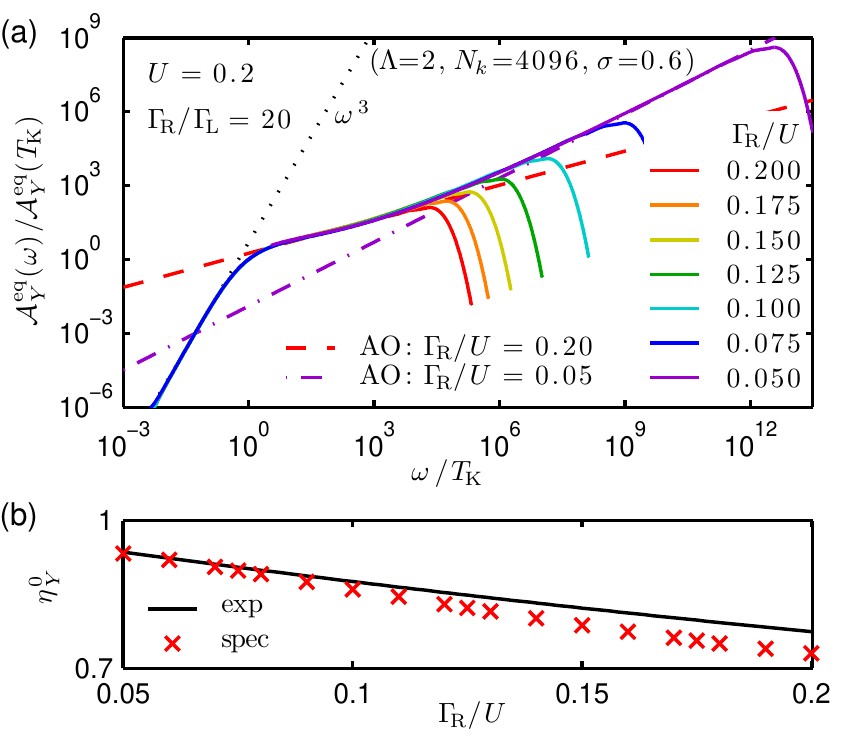}
\par
\end{centering}
\caption{\label{fig:PS_A_Spectra} (Color online)
(a) The pseudospin-flip spectral function $\A^{\equil}_{Y} (\omega)$ 
[cf.\ \eref{eq:opY}] for the PS model without charge sensor, for several 
values of $\GammaR / U$ with fixed ratio $\GammaR / \GammaL$, 
calculated at $\epsd = -U/2$: plotting $\A^\equil_{Y} (\omega)/ 
\A^\equil_{Y} (\TK)$ versus $\omega/\TK$ yields a scaling collapse. 
The frequency dependence of the curves qualitatively changes at $\TK$: 
for $\omega < \TK$ we find Fermi liquid behaviour, $\sim \omega^3$ 
(dotted line), while for $\omega > \TK$ each curve shows a nontrivial 
AO power-law, $\sim \omega^{-1 + 2 \dYp}$ [\eref{eq:spectrumY_AO}], 
exemplified by the dashed and dash-dotted lines for $\GammaR / U 
= 0.2$ and $0.05$, respectively.
(b) Comparison of the values for $\dYp$ expected from \eref{eq:dxy0} 
(solid line), or extracted from the spectral function $\A_Y^\equil (\omega)$ 
by using \Eq{eq:etaY}
in the intermediate-frequency regime between $\TK$ and the high frequency 
maximum (crosses). The accuracy of this extraction deteriorates with increasing 
$\GammaR/U$, since this reduces the width of the latter regime [see (a)], 
causing the relative error between crosses and solid line to increases from 
$1\%$ for $\GammaR / U = 0.05$ to $10\%$ for $\GammaR / U = 0.2$.
}
\end{figure}

Let us now explore the role of AO in population switching. To this
end, we note that the effective exchange interaction $\hat H_{\rm
  exch}$ of \eref{eq:KondoHamilton} is similar in structure to the
IRLM of \eref{eq:SIRL_B_H}: both involve two charge configurations
(the former (0,1) and (1,0), the latter 0 and 1), which induce
different phase shifts in the leads due to a dot-lead interaction
term, and which are connected by a tunnelling term. More formally, 
the relation between the IRLM and PS is revealed by the equivalence 
of both models to the Kondo model (for the IRLM, this equivalence 
is discussed, e.g., in Refs.~\onlinecite{Gogolin,Schlottmann1980, Borda2008}). 
Thus, we may expect AO to play a similar role for both models, and 
hence perform an analysis similar to that in Sections~\ref{sec:IRLM_B} 
and \ref{sec:IRLM_C}.

Specifically, let us study the spectral function $\A^{\equil}_{Y} (\omega)$ 
of the pseudospin-flip operators occurring in 
$\hat H_{\rm exch}$,
\begin{equation}
	\hat{Y}^{\dagger} = 
	\hat{c}_{\rm L}^{\pdag} \hat{c}_{\rm R}^{\dagger} \hat{d}_{\rm L}^{\dagger} \hat{d}_{\rm R}^{\pdag}
	\komma \qquad 
	\hat{Y} = 
	\hat{d}_{\rm R}^{\dagger} \hat{d}_{\rm L}^{\pdag} \hat{c}_{\rm R}^{\pdag} \hat{c}_{\rm L}^{\dagger}
	\punkt \label{eq:opY}
\end{equation} 
These induce transitions between the local charge configurations (0,1)
and (1,0) and simultaneously add an electron to one lead while
removing an electron from the other. (Such a transition does not
constitute a quench, since for the present model $n_{d}$ is not
conserved.) $\A^{\equil}_{Y}$ should, in some respects, be analogous
to $\A_{dc^\dagger}$ of \sref{sec:IRLM_B}.  We have thus calculated
$\A^{\equil}_{Y}$ numerically, using the Hamiltonian $\hat{H}_{\rm
  SIAM}$ of \Eq{eq:SIAM}. Indeed, \fref{fig:PS_A_Spectra}(a), which
shows $\A^\equil_Y(\omega)$ for several values of $\GammaR / U$,
exhibits several features reminiscent of \fref{fig:AOC02spec}(c) for
$\A_{dc^\dagger}(\omega)$: (i) A crossover scale $\ostar$, separating
a regime of very low frequencies from one of intermediate frequencies,
is clearly discernible; it is given by $\ostar \simeq \TK$. (ii) When
properly rescaled by plotting $\A^\equil_Y(\omega) /
\A^\equil_Y(\ostar)$ versus $\omega/\ostar$, all curves collapse onto
each other in the regime $\omega \lesssim D$. (iii) In the
low-frequency regime $\omega \ll \ostar$ we find the same Fermi-liquid
power law, $\A^\equil_Y(\omega) \sim \omega^{3}$ (dotted line), as for
$\A_{dc^\dagger}(\omega)$ [cf.\ \eref{eq:SIRL_B_FL-cddagger}].  (An
analytical explanation for this fact is given in at the end of the
Appendix.) (iv) In an intermediate-frequency regime $\ostar \lesssim
\omega \lesssim \omegahe$, whose upper limit $\omegahe$ is a
high-energy scale set by the minimum of the bandwidth or the cost of
local charge fluctuations, we find an AO-dominated power law,
\begin{subequations}
\label{eq:spectrumY_AO}
\begin{eqnarray}
	\A^\equil_{Y} (\omega) & \sim & \omega^{-1 + 2 \dYp} , \quad
	\ostar \lesssim \omega \lesssim \omegahe \punkt
	 \label{eq:etaY}
\end{eqnarray} 
Though the numerical calculation of $\A^\equil_{Y} (\omega)$ was
performed using the full Hamiltonian $\hat{H}_{\rm SIAM}$ of
\eref{eq:SIAM}, tunnelling is not important on the short time-scales
that govern the frequency regime $\omega > \ostar$. Hence, we expect
the exponent $\dYp$ found from \eref{eq:etaY} to be equal in value to
that which one would obtain in the $\omega \to 0 $ limit of a
calculation performed in the \emph{absence} of pseudospin-flips, 
i.e.\ using $\hat{H}_{\rm exch}$ without the pseudospin-flip terms 
in the second line of \eref{eq:KondoHamilton} (but retaining its first 
and third lines). 

Without pseudospin-flips, the correlator involving $\hat{Y}$ would 
actually constitute a type 2 quench correlator, because $\hat{Y}^{\dagger}$ 
changes $( \hat{n}_{\rm L} - \hat{n}_{\rm R} )$, which is a conserved 
quantum number for Hamiltonians without pseudospin-flips. Therefore, the 
expected value of $\dYp$ can be predicted using the generalized Hopfield 
rule [\eref{eq:Hopfield_dY}]. For the present case of two channels that are 
not interconnected by tunnelling, so that the total charge within each 
channel is conserved, it can be applied to each channel separately, 
adding the corresponding exponents \cite{Weichselbaum2011} [cf.\ 
\Eq{eq:multichannel}]: 
\begin{equation}
	\dYp = \tfrac{1}{2}  ( \dtL + 1 )^2 + \tfrac{1}{2}
 ( -\dtR - 1 )^2 
	\punkt \label{eq:dxy0}
\end{equation}
\end{subequations}
Here $\dtm$ describes the change in phase shift, divided by $\pi$,
induced in lead $\mu$ by a pseudospin-flip; it is given by
\eref{eq:delta_analytical_ph}, with $U$ replaced by $U_{\!\mu}$ [from
\eref{eq:U_alpha}]. The applicability of these arguments is confirmed
by \fref{fig:PS_A_Spectra}(b), which shows that the exponents
extracted from the numerical spectra (crosses) agree quite well with
the values expected from \eref{eq:dxy0} (solid line). The quality of
the agreement deteriorates with increasing $\GammaR / U$, because this
reduces the width of the intermediate-frequency regime, making an
accurate extraction of $\dYp$ from $\A_Y^\equil(\omega)$ increasingly
difficult.

\Eref{eq:dxy0} allows us to understand why PS is always continuous in 
this model: Since $-1 \leq \dtm \leq 0$, the scaling dimension of $\hat{Y}$ 
satisfies $\dY \leq 1$, implying that this operator always remains a 
relevant perturbation that does not flow to zero at low energy scales. 
This means that AO, although present, is not strong enough to completely 
suppress the amplitude for pseudospin-flip transitions. Hence, the two 
sectors (0,1) and (1,0) are always coupled by the effective low-energy 
Hamiltonian, so that population switching is continuous.\cite{Goldstein2010,Goldstein2011} 

%---------------------------------------------------------------------
\subsection{AO induced by quench of level positions}\label{sec:quench-nosensor}
%---------------------------------------------------------------------

\begin{figure}
\begin{centering}
\includegraphics{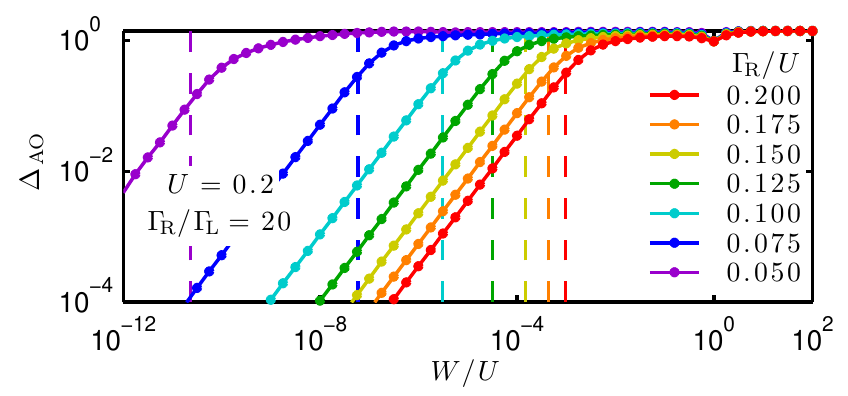}
\par
\end{centering}
\caption{\label{fig:PS_A_Sweep} (Color online)
AO for the PS model without charge sensing. The exponent $\dt$ [extracted 
from \eref{eq:Intro_overlap}] (solid lines with dots) is shown as function of 
quench size $W/U$ [\eref{eq:PS_A_quench-by-W}], for several values of 
$\GammaR/U$ with fixed ratio $\GammaR / \GammaL$, showing that AO 
becomes significant once $W$ increases past $\TK / b_{0}$ (indicated by 
dashed vertical lines). The exponent $\dt$ increases linearly with $W$ for 
$W \ll \TK / b_{0}$, and saturates to a maximal value of $\sqrt{2}$ 
[\eref{eq:PS_A_dAOmax}] for $W \gg \TK / b_{0}$. The corresponding 
values of $\dch$ [from \Eq{eq:Intro_Friedel}] are not shown but agree 
with $\dt$ with relative errors of a few percent.
}
\end{figure}

As mentioned above, the operators $\hat{Y}^{\dagger}$ and $\hat{Y}$ 
connect two configurations with different local occupancies, (0,1) and 
(1,0). To shed further light on the AO between such configurations, we 
now perform a quantum quench analysis similar to that of \sref{sec:IRLM_C}. 
We consider a type 1 quench, $\Hi \to \Hf$, induced by changing the 
level position $\epsd$ from a value above the symmetry point, favouring 
(0,1), to one below, favouring (1,0): 
\begin{eqnarray}
	\epsd^{\initial} = -U/2 + W/2
	\quad \stackrel{{\rm quench}}{\longrightarrow} \quad
	\epsd^{\final} = -U/2 - W/2
	\punkt \label{eq:PS_A_quench-by-W}
\end{eqnarray}
The corresponding ground states, $\ket{G_{\initial}}$ and $\Gf$, will 
display AO as in \Eq{eq:Intro_overlap}. Based on the lessons learnt from 
\sref{sec:IRLM_C}, the corresponding exponent $\dt$ will increase with 
the width $W$ of the quench. Indeed, \fref{fig:PS_A_Sweep} [to be compared 
with \fref{fig:AOC03_Sweep}(a)] shows that $\dt$ increases from close to 
$0$ for $W$ much below $\TK / b_{0}$ (indicated by vertical dashed lines) 
to a maximal value of 
\begin{eqnarray}
	\dt^{\rm max} = \sqrt{(1)^2 + (1)^2} = \sqrt{2}
 \label{eq:PS_A_dAOmax}
\end{eqnarray}
for $W \gg \TK / b_{0}$. This maximal value reflects the displaced 
charge $\dch$ [cf.\ \Eq{eq:Intro_Friedel}] induced by a very strong 
quench: both $n_{\rm L}$ and $n_{\rm R}$ are $\simeq 0$ (or 
$\simeq 1$) if the level position is far above (or below) the Fermi 
energy, $\epsd^{\initial} = -U/2 + W/2 \gg 0$ (or $\ll 0$), cf.\
\frefsub{fig:PS_A}{b}, thus the displaced charge associated with both 
$n_{\rm L}$ and $n_{\rm R}$ is $1$. (The contribution to $\dch$ from 
the leads turns out to be negligible here, \cite{Weichselbaum2011} since 
for sufficiently large $W$ the Fermi sea is essentially decoupled from the 
dot.)

%---------------------------------------------------------------------
\subsection{Summary for  PS without sensor}\label{sec:results-nosensor}
%---------------------------------------------------------------------

The results of this section can be summarized as follows: (i) The
energy scale setting the width of PS is proportional to $\TK$. (ii) 
This can directly be attributed to AO: as shown in 
\fref{fig:PS_A_Sweep}, the ground states of two configurations on
opposite sides of the switching points exhibit strong AO when their
level positions differ by more than $\TK / b_{0}$. Thus, quantum 
fluctuations between them, induced by operators such as $\hat{Y}$ 
and $\hat{Y}^{\dagger}$, are strongly suppressed. (iii) For the present
model PS will always be continuous as a function of $\epsd$, because
(for given $U$) $\TK$ is nonzero for any fixed choice of $\GammaL$ 
and $\GammaR$ (although exponentially small), and AO ceases to be 
important ($\dt \simeq 0$) once $\epsd$ comes within $\TK / b_{0}$ 
of the switching point. Conversely, however, it should now also be 
plausible that an essentially abrupt PS will be achievable if, by a 
suitable modification of the model, the degree of AO between the
configurations (0,1) and (1,0) can be enhanced sufficiently to push
$\TK$ to zero even for finite $\GammaL$ and $\GammaR$. As pointed out
by GBG, \cite{Goldstein2010,Goldstein2011} this can be achieved by
adding a charge sensor, to which we turn next.

%---------------------------------------------------------------------
\section{Population Switching with Sensor}\label{sec:PS_C}
%---------------------------------------------------------------------

\begin{figure}[h]
\begin{centering}
\includegraphics{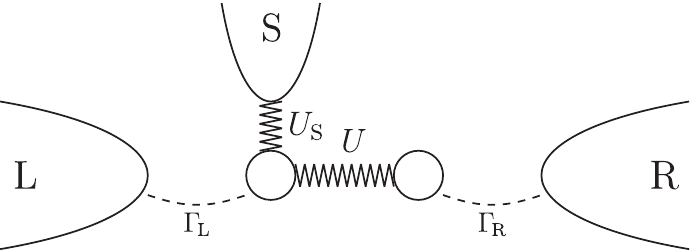}
\par
\end{centering}
\caption{\label{fig:PS_C}
Cartoon of the Hamiltonian (\ref{eq:H_GBG}), describing an 
asymmetric SIAM with an additional sensor lead coupled 
electrostatically to the left dot.
}
\end{figure}

In this section we study the effects of adding an electrostatically coupled 
charge sensor to the model of the previous section, as proposed by GBG, 
\cite{Goldstein2010,Goldstein2011} and analyse how this enhances the effects of AO. In 
particular, we show that by increasing the sensor coupling strength ($\US$), 
the effective Kondo temperature ($\TKS$) can be driven to zero, implying 
that population switching becomes abrupt. (A study of how additional leads 
increase the effects of AO for static quantities has recently been performed 
in similar context, involving a multi-lead IRLM. \cite{Borda2008} )

%---------------------------------------------------------------------
\subsection{Width of switching regime}\label{sec:PSwidth_B}
%---------------------------------------------------------------------

\begin{figure}
\begin{centering}
\includegraphics{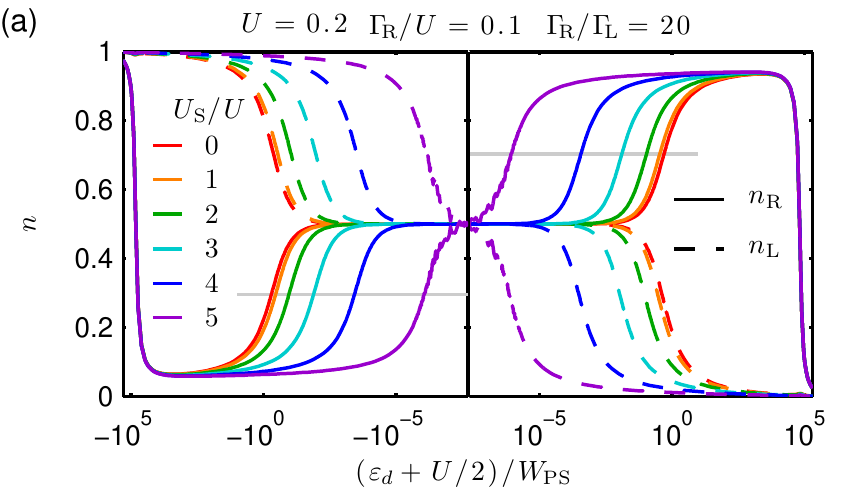}
\includegraphics{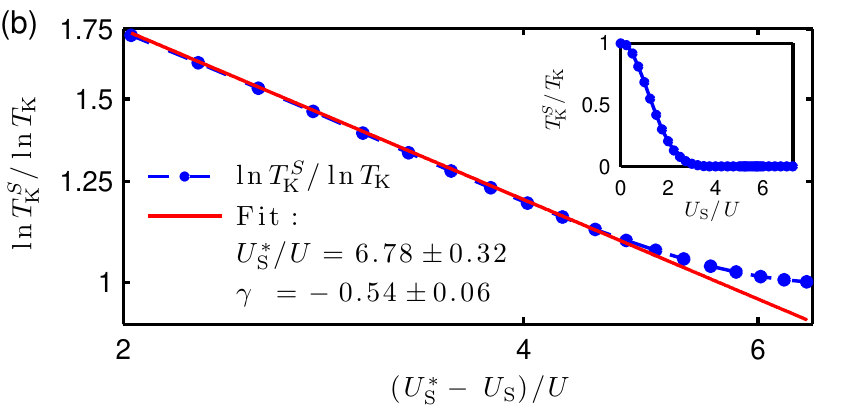}
\par
\end{centering}
\caption{\label{fig:PS_C_TKS} (Color online) Population switching for
  the charge sensor model of \eref{eq:H_GBG}.  (a) $n_{\rm R} (\epsd)$
  (solid lines) and $n_{\rm L} (\epsd)$ (dashed lines) for several
  values of $\US/U$, plotted versus $( \epsd + U/2 ) / \DPS$ in a
  pseudo-logarithmic fashion (\qm{pseudo} in that the x-axis is
  plotted logarithmic with positive and negative values to the left
  and right of the switching point, respectively, represented by the
  vertical solid line). The horizontal light solid lines indicate the
  values of $n_{\rm R}$ which define the widths $\DPSS$ of the PS
  regimes. The noisy behaviour of the curves for $\US = 5U$ at small
  values of $\epsd$ indicates that our analysis cannot resolve smaller
  values for $\epsd$ as we are reaching the limits of double precision
  numerical accuracy.  (b) Inset: $\TKS / \TK$ as function of $\US /
  U$, showing the rapid decrease of the Kondo temperature with
  increasing coupling. Main panel: $\ln \TKS / \ln \TK$ versus
  $(\US^{*} - \US)/U$, plotted on a log-log scale (dashed line with
  dots), together with a linear fit using \eref{eq:TKS_scaling} (solid
  line).  }
\end{figure} 

GBG proposed to extend the asymmetric SIAM studied above by
introducing a third lead as \qm{charge sensor} for the left dot (see
\fref{fig:PS_C}). For simplicity, it is taken to have the same density 
of states as the other two leads, but in contrast to the latter, it couples 
to the left dot only electrostatically (\emph{not} by tunnelling), with 
interaction strength $\US$ (with $\hat{c}_{\rm S} \equiv \sum_{\varepsilon} 
\hat{c}_{\varepsilon \rm S}$):
\begin{eqnarray}
	\hat{H} & = & \hat{H}_{\rm SIAM} 
	\label{eq:H_GBG} \\
	& & + \sum_{\varepsilon} \varepsilon \, \hat{c}_{\varepsilon \rm S}^{\dagger} \hat{c}_{\varepsilon \rm S}^{\pdag}
	+ \US
	( \hat{d}_{\rm L}^{\dagger} \hat{d}_{\rm L}^{\pdag} - \frac{1}{2} )
	( \hat{c}_{\rm S}^{\dagger} \hat{c}_{\rm S}^{\pdag} - \frac{1}{2} )
	\punkt \nonumber
\end{eqnarray}
A plot of $n_{\rm L}$ and $n_{\rm R}$ as functions of $\epsd$ for this model
looks essentially similar to \fref{fig:PS_A}(b), showing population
switching at $\epsd = - U/2$. However, when the strength of the
coupling $\US$ is increased, the width of the PS, say $\DPSS$, is
strongly reduced below the value $\DPS$ it had for $\US = 0$, as
predicted by GBG. This is illustrated in \fref{fig:PS_C_TKS}(a), which
shows $n_{\rm R}$ (solid lines) and $n_{\rm L}$ (dashed lines) as functions of
$(\epsd + U/2) / \DPS$, using a logarithmic scale to zoom in on the
immediate vicinity of the PS. In fact, as $\US$ approaches a critical
value $\US^{*}$, the width $\DPSS$ drops exponentially towards zero,
until it becomes too small to be resolved within double precision 
numerical accuracy.

The behaviour of $\DPSS$ is mimicked by that of the Kondo temperature, 
calculated via the pseudospin susceptibility, $\TKS \equiv 1/\chi_s$. We 
find that it decreases relative to its $\US = 0$ value $\TK$, precisely in 
proportion to $\DPSS$, such that 
\begin{equation}
	\frac{\TKS}{\TK} = \frac{\DPSS}{\DPS}
	\label{eq:TKS}
\end{equation}
holds within our numerical accuracy. 

The transition from a continuous to an abrupt PS as $\US$ crosses 
$\US^{\ast}$ has been predicted to be of the Kosterlitz-Thouless type. 
\cite{Goldstein2010,Goldstein2011} This implies that $\TKS$ is expected to approach 
zero according to 
\begin{equation}
	- \ln \TKS \sim \left( \US^{*} - \US \right)^{\gamma}
	\komma \label{eq:TKS_scaling}
\end{equation}
where $\gamma = - 1/2$. To test whether our data is conform to this
expectation, \fref{fig:PS_C_TKS}(b) shows $\ln(\TKS)/\ln(\TK)$ vs.\
$(\US^{\ast} - U)$ on a log-log plot.  Indeed, we find a straight line
for $\US$ not too close to $\US^{\ast}$, consistent with
\eref{eq:TKS_scaling}. We extract the values $\gamma = - 0.54 \pm
0.06$ and $\US^{\ast} / U = 6.78 \pm 0.32$, by making linear fits over
several somewhat different fitting ranges and taking the average and
standard deviation of the fit parameters as final fitting results. The
relatively large errors of about $10\%$ are a consequence of the fact
that it is not possible to obtain data for $\US$ closer to $\US^{*}$,
since this would drive $\TKS$ below the level of numerical noise. 

We note that analytical calculations based on Refs.~\onlinecite{Goldstein2010} 
and \onlinecite{Goldstein2011} [using the more accurate criterion, $J_{z} (\US^{\ast}) 
= J_{xy} (\US^{\ast})$ in the notation of these papers] predict the critical interaction 
to be $\US^{\ast} / U \sim 7.6$. The agreement of this prediction with the numerical 
result of $6.8$ is quite respectable, given the inaccuracies in both the numerical and 
analytical calculations [for the latter, inaccuracies arise since the cutoff scheme 
employed in the analytical calculation is different from the one realized numerically. 
The cutoff appears explicitly in the arguments of the functions $Q$ in Eqs.~(6) to (10) 
of Ref.~\onlinecite{Goldstein2010}]. 

Though the above results unambiguously show that the width of PS
decreases exponentially as $\US$ approaches a critical value
$\US^{\ast}$, an analysis based purely on $\DPSS$ can not access the
critical point itself or the regime beyond. We therefore proceed now
with {a numerical calculation of the dynamics of the pseudospin-flip
operator, for which we are not constrained to $\US < \US^{\ast}$.

%---------------------------------------------------------------------
\subsection{AO in dynamics of pseudospin-flip
operator}\label{sec:AO-AY-withsensor}
%---------------------------------------------------------------------

\begin{figure}
\begin{centering}
\includegraphics{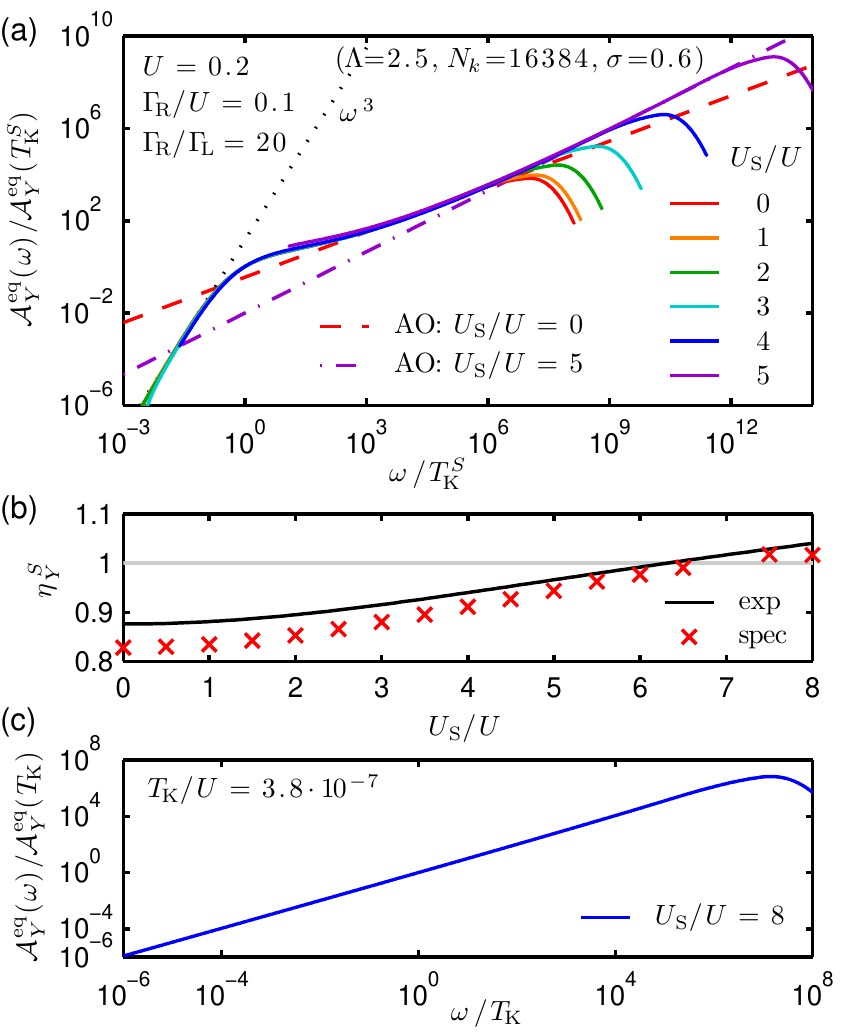}
\par
\end{centering}
\caption{\label{fig:PS_C_Spectra} (Color online) (a) The
pseudospin-flip spectral function $\A_{Y}^\equil (\omega)$ [cf.\
\eref{eq:opY}] for the PS model with charge sensor, for several values
of $\US / U$, calculated at $\epsd = -U/2$: plotting $\A^\equil_{Y}
(\omega) / \A^\equil_{Y} (\TKS)$ versus $\omega/\TKS$ yields a scaling
collapse.  The general shape of the curves is similar to those shown
in \fref{fig:PS_A_Spectra}: for $\omega < \TKS$ we find Fermi liquid
behaviour, $\sim \omega^3$ (dotted line), while for $\omega > \TKS$
each curve shows a nontrivial AO power-law, $\sim \omega^{-1 + 2
\dYS}$ [cf.\ \eref{eq:spectrumY_AO}], exemplified by the dashed and
dash-dotted lines for $\US / U = 0$ and $5$, respectively.  (b)
Comparison of the values for $\dYS$ expected from \eref{eq:dYS} (solid
line), or extracted from the spectral function $\A_Y^\equil(\omega)$
in the intermediate-frequency regime between $\TK$ and the high
frequency maximum (crosses). The relative errors are below $5\%$,
where the errors decrease with increasing $\US$ for similar reasons as
in \fref{fig:PS_A_Spectra}.  The light horizontal line indicates $\dYS
= 1$. (We were unable to obtain reliable data for $\US$ around $7U$,
presumably because this is too close to $\US^{\ast}$.)  (c)
$\A_Y^\equil(\omega) / \A_Y^\equil(\TK)$ versus $\omega / \TK$ for
$\US = 8U$. The AO power-law behaviour $\omega^{-1 + 2 \dY^S}$ extends
down to the smallest frequencies accessible, illustrating that the
crossover scale $\TKS$ has become undetectably small.  }
\end{figure}

The reason for the $\US$-dependence of $\DPS$ and $\TKS$ is that the
introduction of the sensor ($\US \neq 0$) increases the influence of
AO in the leads. As pointed out by GBG, the scaling dimension of
$\hat{Y}$ acquires an extra contribution $\tfrac{1}{2} \dtS^2$ due to
the sensor lead:
\begin{equation} \dYS = \tfrac{1}{2} ( \dtL + 1 )^2 + \tfrac{1}{2}(
-\dtR - 1 )^2 + \tfrac{1}{2} \dtS^2 \komma \label{eq:dYS}
\end{equation} where $\dtS$ is given by \eref{eq:delta_analytical_ph},
with $\US$ replacing $U$. By increasing $\US$ and thereby $\dtS^2$, it
is thus possible to drive $\dYS$ beyond $1$. This will render the
pseudospin-flip operators $\hat{Y}$ and $\hat{Y}^{\dagger}$
\emph{irrelevant}, thus suppressing quantum fluctuations between the
(0,1) and (1,0) configurations, and, hence, pushing $\TKS$ down to zero. 

To check this scenario explicitly, we have studied the $\US$-dependence of 
$\dYS$ by extracting it from the spectral function $\A^\equil_{Y}(\omega)$, 
calculated at the particle-hole symmetric point for several values of $\US$. 
The general shape of $\A^\equil_{Y}$, shown in \frefsub{fig:PS_C_Spectra}{a}, 
is similar to that of \frefsub{fig:PS_A_Spectra}{a} for $\US=0$: For frequencies 
well below $\TKS$, $\A^\equil_{Y}(\omega)$ scales as $\omega^3$, while in 
the regime of intermediate frequencies, $\TKS \lesssim \omega \lesssim \omegahe$ 
(cf.\ \sref{sec:AO-AY-withoutsensor}), the spectrum shows AO power-law 
behaviour, $\sim \omega^{-1 + 2 \dYS}$. Indeed, \frefsub{fig:PS_C_Spectra}{b} 
shows that the values for $\dYS$ extracted from the spectra (crosses) agree 
fairly well with those expected from \eref{eq:dYS}.  Moreover, for sufficiently 
large $\US / U$, the exponents $\dYS$ increase past 1, confirming that the
pseudospin-flip operators become irrelevant.

%---------------------------------------------------------------------
\subsection{AO induced by quench of level positions} \label{sec:quench-withsensor}
%---------------------------------------------------------------------

\begin{figure}
\begin{centering}
\includegraphics{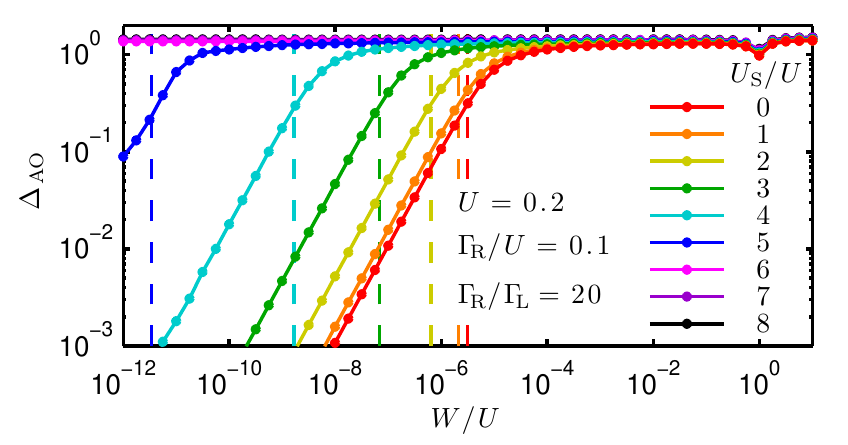}
\par
\end{centering}
\caption{\label{fig:PS_C_Sweep} (Color online)
AO for the PS model with charge sensing. The exponent $\dt$ [extracted 
from \eref{eq:Intro_overlap}] (solid lines with dots) is shown as function 
of quench size $W/U$ [\eref{eq:PS_A_quench-by-W}], for several values 
of $\US/U$, with fixed values of $\GammaR$ and $\GammaL$. We see 
that $\TKS / b_{0}$ (indicated by dashed vertical lines) is pushed to zero 
as $\US$ increases past $\US^{\ast} / U \simeq 6.78$. Already for $\US / U 
\geq 6$ the curves are essentially indistinguishable, in that they do not 
deviate from their constant  value for all $W/U$-values accessible to our 
analysis. For $W \gg \TKS / b_{0}$ the exponent $\dt$ saturates to a 
maximal value given by \eref{eq:PS_C_dAOmax}. The corresponding values 
of $\dch$ [from \eref{eq:Intro_Friedel}] are not shown but agree with $\dt$ 
with relative errors of a few percent. 
}
\end{figure}

To further highlight the effect of AO on $\TKS$, let us consider again 
the quench of level position [\eref{eq:PS_A_quench-by-W}] studied in 
\sref{sec:quench-nosensor}, and repeat the analysis presented there, 
but now for several different values of $\US / U$. \Fref{fig:PS_C_Sweep} 
shows the results for the exponent $\dt$. For large values of $W$ the 
AO factor reaches its maximal value 
\begin{eqnarray}
	\dt^{\rm max} = \sqrt{(1)^2 + (1)^2 + \dtS^2 }
	\punkt \label{eq:PS_C_dAOmax}
\end{eqnarray}
This is similar to \eref{eq:PS_A_dAOmax} for the model without 
sensor, but includes the additional contribution $\dtS^2$ [given by 
\eref{eq:delta_analytical_ph}, with $\US$ replacing $U$] from 
the displaced charge induced in the sensor lead by the change in 
local occupancy of the left dot from $n_{\rm L} = 0$ to $1$. 

The most important feature of \fref{fig:PS_C_Sweep} is the fact that 
the crossover scale $\TKS / b_{0}$ (indicated by vertical dashed lines) 
is rapidly pushed to extremely small values as $\US / U$ is increased. 
Indeed, for $\US = 8 U$, which lies beyond the critical value of 
$\US^{*} /U \simeq 6.78$ discussed above, $\dt$ is essentially 
pinned to its maximal value down to the smallest values of quench 
range $W$ that we can access numerically. This is consistent with 
the fact that the corresponding spectral function $\A_Y^\equil(\omega)$ 
at $\US = 8 U$, shown in \fref{fig:PS_C_Spectra}(c), shows nontrivial 
AO power laws down to the lowest frequencies accessible, with no 
trace of a Fermi-liquid $\omega^3$. This demonstrates very clearly, 
if somewhat indirectly, that the PS will be abrupt for $\US > \US^{\ast}$. 

%---------------------------------------------------------------------
\subsection{Summary for  PS with sensor}\label{sec:results-withsensor}
%---------------------------------------------------------------------

Let us  summarize the results of this section, by way of extending the 
list of salient points collected in \sref{sec:results-nosensor}. (iv) The 
presence of a charge sensor reduces the crossover scale $\TKS$, which 
reaches zero at a critical coupling $\US^{\ast}$ [\fref{fig:PS_C_TKS}]. (v) 
This reduction is due to the increased effect of AO in the leads, which 
increases the scaling dimension $\dYS$ [\fref{fig:PS_C_Spectra}]; 
when the latter passes 1 (corresponding to $\US = \US^{\ast}$), the 
pseudospin-flip operators become irrelevant and $\TKS$ equals zero, 
rendering the PS abrupt. (vi) Correspondingly, for $\US > \US^{\ast}$, the 
spectrum $\A_Y^\equil(\omega)$ shows nontrivial AO power-law behaviour, 
$\omega^{-1 + 2 \dYS}$, all the way down to the smallest frequencies 
accessible [\frefsub{fig:PS_C_Spectra}{c}], and a low-frequency regime 
showing Fermi-liquid exponents does not exist. 
 
%---------------------------------------------------------------------
\section{Concluding remarks}\label{sec:Conclusions}
%---------------------------------------------------------------------

The goal of this paper was to elucidate the role of the Anderson
orthogonality catastrophe in giving rise to anomalous scaling
dimensions in dynamical correlation functions for quantum impurity
models. To this end, we have studied several setups involving
(interacting) quantum dots and (non-interacting) leads. The quantum
dots and leads may be interconnected electrostatically, or also
through tunnel-coupling. In our analysis we focussed on the asymptotic
behaviour of various correlation functions $\G(t)$ and the
corresponding spectral functions $\A(\omega)$ in the limit of long
times or low frequencies, respectively. Their asymptotic behaviour
could be understood via a generalized version of Hopfield's rule,
whose validity was checked and confirmed through an extensive NRG
analysis. As a particular application, we performed a detailed study
of population switching, both without and with a third lead that acts
as a charge sensor. We confirmed a previous prediction
\cite{Goldstein2010,Goldstein2011} that when the charge sensor is
sufficiently strongly coupled, population switching can turn into an
abrupt quantum phase transition.

Aside from presenting a systematic discussion of the generalized Hopfield 
rule, which, hopefully, will be useful for practitioners in the fields, several 
general features have emerged from our analysis:

\begin{figure}
\begin{centering}
\includegraphics{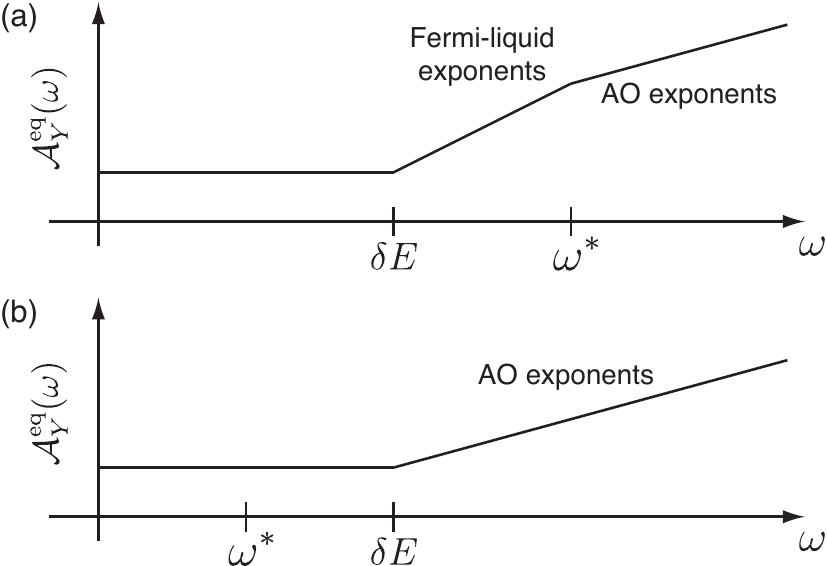}
\par
\end{centering}
\caption{\label{fig:schematic-intermediate-regime} 
Schematic depiction of an equilibrium spectral function
$\A_Y^\equil(\omega)$ for the cases that the local 
charge relaxation rate $\ostar$ is (a) larger or (b) 
smaller than the lead level spacing $\delta E$.
}
\end{figure}

(1) In the context of a local quantum quench of type 1, where a change
of parameters switches the Hamiltonian from $\Hi$ to $\Hf$, each
lead-dot electrostatic coupling gives rise to an AO factor in the
ground state overlap $\vert \braket{G_{\initial}}{G_{\final}} \vert$,
reflecting a change in the many-body configuration of the lead when
the charging state of the dot is modified. This AO factor scales as
$N_{\mu}^{-\frac{1}{2} \dtm^2}$, where $N_{\mu}$ is the number of
electrons in lead $\mu$ and $\dtm$ the change in the scattering phase,
divided by $\pi$, in that lead. (AO factors from leads that are
not interconnected by tunnelling, so that the total charge within each
channel is conserved, are multiplicative
[\Eq{eq:multichannel}].\cite{Weichselbaum2011} )

(2) AO also arises for a type 2 quench, induced by an operator 
$\hat{X}^{\dagger}$ that connects initial and final ground states 
$\Gi$ and $\Gf$ lying in dynamically disconnected sectors of Hilbert 
space. In particular, AO influences the corresponding quench spectral 
function $\A_{X} (\omega)$ which scales as $\A_X ( \omega ) \sim 
\omega^{-1 + \dtXsq}$ [\eref{eq:Intro_A_X}]. For a Hamiltonian without 
tunnelling terms such as the LCM of \eref{eq:LCM}, the spectral function 
for $\hat{X}^{\dagger} = \hat{d}^{\dagger}$ thus scales as ${\A_{d} ( \omega ) 
\sim \omega^{-1 + \dtd^2}}$. 

(3) When a type 2 quench has the form of a tunnelling operator,
$\hat{Y}^\dagger = \hat{c} \hat{d}^{\dagger}$, the asymptotic power
law is modified to become $\A_{dc^{\dagger}} \sim \omega^{-1 + (\dtd +
  1)^2}$ [\Eq{eq:SIRL_A_Adct}], implying a scaling dimension $\ddctp =
\tfrac{1}{2} (\dtd + 1)^2$. For a particle-hole symmetric interaction
term [as in \Eq{eq:SIRL_B_H}], we have $-1 \le \dtd \le 0$
[\Eq{eq:delta_analytical_ph}], implying that $0 \le \ddctp \le 1/2$,
thus tunnelling between a dot and a single lead is always a relevant
perturbation.

(4) The scaling exponent can be increased, and AO strengthened, by
coupling the dot(s) to further leads. In particular, leads that couple
to the dot only electrostatically (not via tunnelling) contribute AO
exponents of the form $\frac{1}{2} \dtm^2$, and thus enhance AO 
more strongly than leads that are tunnel-coupled [cf. point (3)]. In 
this way, the scaling dimension of the tunnelling operator can be 
increased past 1 [cf.\ \Eq{eq:dYS}], and tunnelling rendered irrelevant. 
In such a situation, population switching becomes a quantum phase 
transition, tuned by gate voltage. 

(5) A particularly revealing way of demonstrating the effect of AO for
population switching is to calculate the exponent $\dt$ for a type 1
quench in which the level position is abruptly changed from lying
above to below the PS point (see Figs.~\ref{fig:PS_A_Sweep} and
\ref{fig:PS_C_Sweep}, which are analogous to \fref{fig:AOC03_Sweep}(a)
for the IRLM).

(6) In the presence of tunnelling terms of the form $(\hat{c}^{\dagger} 
\hat{d} + \hat{d}^{\dagger} \hat{c})$, operators such as $\hat{Y}^{\dagger} 
= \hat{d}^{\dagger}$, $\hat{c} \hat{d}^{\dagger}$ and $\hat{c}^{\dagger} 
\hat{d}^{\dagger}$ do not induce a quench, since they do not cause a 
switch between disconnected sectors of Hilbert space. Thus, when such 
an operator acts on the ground state, the resulting state will relax back 
to the ground state over long time scales, say $t \gg 1/\ostar$, where 
$\ostar$ represents the local charge relaxation rate.

(7) The corresponding equilibrium spectral function
$\A_Y^\equil(\omega)$ thus typically shows trivial Fermi-liquid
exponents [e.g.\ \eref{eq:SIRL_B_FL}] in the regime of very small
frequencies, $\delta E \lesssim \omega \ll \ostar$, where $\delta E$
represents an infrared cutoff such as the level spacing in the
lead. (Throughout this paper we took $\delta E \simeq 0$, since in NRG
calculations $\delta E$ can be made arbitrarily small by using
sufficiently long Wilson chains.)

(8) In an intermediate frequency regime $\omega > \ostar$, the
equilibrium spectral function $\A_{Y}^{\equil} (\omega)$ may
nevertheless contain signatures of anomalous AO exponents, scaling as
$\omega^{-1 + 2 \etap_Y}$ [e.g.\ \eref{eq:SIRL_B_AOspec}], where
$\etap_Y$ represents the scaling dimension of $\hat{Y}$ calculated in
the absence of tunnelling. Thus, such exponents may be extracted by
focussing on this regime of intermediate frequencies [as done in
\frefs{fig:AOC02spec}, \ref{fig:PS_A_Spectra} and \ref{fig:PS_C_Spectra}]. 
This is schematically indicated in \frefsub{fig:schematic-intermediate-regime}{a}.

(9) If AO can be made so strong that the scaling dimension $\eta_Y^0$
of the operator $\hat{Y}^\dagger$ is larger than 1, the scale $\ostar$ is
pushed below $\delta E$ (or, in the context of NRG, below the level of
numerical noise). In this case, the regime of anomalous AO scaling
$\omega^{-1 + 2 \etap_Y}$ will extend all the way down to the smallest
frequencies accessible [e.g. \ref{fig:PS_C_Spectra}(c)], as
schematically indicated in
\fref{fig:schematic-intermediate-regime}(b).

To conclude, we note that cases where AO dominates in the 
low frequency limit such that $\ostar \simeq 0$, [as in point (9)], 
quantum fluctuations of the charge on the dot(s) are essentially 
completely frozen out. At zero temperature and in the absence of 
any extraneous decay mechanism, the system will remain localized 
in a particular local charge configuration. Thus, varying the gate 
voltage in such a situation may lead to hysteretic behaviour. It would 
be very interesting to experimentally search for such signatures of 
the freezing out of charge fluctuations by performing linear response 
measurements at the PS point.

%---------------------------------------------------------------------
\begin{acknowledgments}
We thank R.\ Berkovits, L.\ Borda, Y.\ Imry, Y.\ Oreg, A.\ Schiller, G.\ Zar\'and, 
and A.\ Zawadowski for helpful discussions. This work received support from 
the DFG (SFB 631, De-730/3-2, De-730/4-2, SFB-TR12, WE4819/1-1), in 
part from the NSF under Grant No.\ PHY05-51164, from the Israel-Russia 
MOST grant, the Israel Science Foundation, and the EU grant under the 
STREP program GEOMDISS. Financial support by the Excellence Cluster 
\qm{Nanosystems Initiative Munich (NIM)} is gratefully acknowledged. 
M.G.\ is supported by the Adams Foundation of the Israel Academy of 
Sciences and Humanities, the Simons Foundation, the Fulbright Foundation, 
and the BIKURA (FIRST) program of the Israel Science Foundation.
\end{acknowledgments}

%---------------------------------------------------------------------
\appendix*
\section{Fermi-liquid spectral functions}\label{sec:AppendixFL}
%---------------------------------------------------------------------

In this appendix we study analytically the low energy ($\omega < \ostar$) 
behaviour of the spectral functions of the IRLM (\sref{sec:IRLM_B}) and the 
PS setup (Sections~\ref{sec:AO-AY-withoutsensor} and \ref{sec:AO-AY-withsensor}).

Let us start from the noninteracting resonant level (\eref{eq:SIRL_B_H} with $U=0$). 
In that case an elementary calculation gives for the retarded dot Green function, 
\cite{Mahan}
\begin{eqnarray}
	\G_{d}^{R} (\omega)
	= \frac{1}{\omega - \epsd - \frac{\Gamma}{\pi\rho} \G_{c}^{R,0} (\omega)}
	= \frac{1}{\omega - \epsd + i \Gamma}
	\komma
\end{eqnarray}
where $\G_{c}^{R,0}$ is the retarded $c$ Green function for $\Gamma = 0$, 
and we assumed the wide band limit (used just to simplify expressions, but 
actually not essential for any of the following arguments) $\G_{c}^{R,0} (\omega) 
= -i\pi\rho = -i\pi/2$ in units where $D=1$. The imaginary part of the retarded 
Green function gives (up to a factor of $-1/\pi$) the well-known Lorentzian 
spectral function
\begin{eqnarray}
\label{eq:Lorenztian-spectral-function}
	\A_{d}^{\rm eq} (\omega)
	= \frac{1}{\pi} \frac{\Gamma}{(\omega - \epsd)^2 + \Gamma^2}
	\punkt
\end{eqnarray}
Thus, at low energies ($\omega \ll \Gamma$) $\A_{d}^{\rm eq} (\omega)$
becomes a constant, corresponding to $\eta_{d}^{\rm eq} = 1/2$ [which 
reproduces \eref{eq:SIRL_B_FL-ddagger}]. This behaviour is easy to understand: In
the absence of tunnelling $\A_{c}^{{\rm eq},0} (\omega) = \rho$ is
constant, reflecting the constant local density of states of the lead
electrons near the end of the lead. In the presence of tunnelling, at
low energy the dot level is well-hybridized with the lead, and assumes
the role of the end of the lead, thus featuring the slowly-varying
low-energy spectral function $\A_{d}^{\rm eq} (\omega)$.

Based on similar arguments, one would expect that, in the presence of tunnelling, 
$\A_{c}^{\rm eq} (\omega)$ is still constant at low-energies, since in that limit the 
small spatial separation between the dot and the end of the lead should be unimportant. 
However, commensurability at half filling (particle-hole symmetry) makes things bit 
more complicated. An explicit calculation gives:
\begin{eqnarray}
	\G_{c}^{R} (\omega)
	& = & \G_{c}^{R,0} (\omega) + \G_{c}^{R,0} (\omega) \sqrt{\frac{\Gamma}{\pi\rho}}
	\G_{d}^{R} (\omega) \sqrt{\frac{\Gamma}{\pi\rho}} \G_{c}^{R,0} (\omega)
	\nonumber \\
	& = & -i\pi\rho \frac{\omega - \epsd}{\omega - \epsd + i\Gamma}
	\punkt
\end{eqnarray}
Thus, when $\epsd$ is nonzero, we indeed get a constant low energy
limit, i.e.\ $\eta_{c}^{\rm eq} = 1/2$. However, when $\epsd = 0$ (the
value used throughout this paper for the IRLM), $\G_{c}^{R} (\omega)
\sim \omega$ while $\A_{c}^{\rm eq} (\omega) \sim \omega^2$,
corresponding to $\eta_{c}^{\rm eq} = 3/2$. To understand this
behaviour, let us examine a half infinite tight-binding chain with
lattice spacing $a$ and Hamiltonian $\hat{H}_{\rm TB} =
\sum_{n=1}^{\infty} ( \hat{\Psi}_{n+1}^{\dagger}
\hat{\Psi}_{n}^{\pdag} + \rm{h.c.}  )$. Taking the continuum limit in
the standard way, we can expand the fast-varying annihilation
operators $\Psi_{n}$ in terms of slowly-varying (on the scale of the
Fermi wavelength) right/left moving fields $\psi_{R/L}(x)$, with $x =
n a$:
\begin{eqnarray}
	\Psi_{n} = e^{i \kF n a} \psi_{R} (n a) + e^{-i \kF n a} \psi_{L} (n a)
	\komma
\end{eqnarray}
where $\kF$ is the Fermi wavevector. From the boundary condition $\Psi_{0} = 0$ 
one gets $\psi_{L} (0) = -\psi_{R} (0)$, so we can define the single slowly-varying 
field $\psi (x)$ by $\psi (x) = \psi_{R} (x)$ if $x>0$ and $\psi(x) = -\psi_{L} (-x)$ 
if $x<0$. Then: 
\begin{eqnarray}
	\Psi_{n} = e^{i \kF n a} \psi (n a) - e^{-i \kF n a} \psi (-n a)
	\punkt
\end{eqnarray}
At half filling, $\kF a = \pi/2$, we get at the site next to the boundary 
\begin{eqnarray}
 	\Psi_{n=2} = - \psi (2 a) + \psi (-2 a) \sim -4 a \partial_{x} \psi (0)
 	\komma
\end{eqnarray}
The same thing happens at the \emph{first site ($n=1$) when we attach
  a dot, since at low energies the dot behaves as the new first
  site}. The spatial derivative is equivalent to a time derivative, up
to the Fermi velocity $v_{\rm F}$. This extra time derivative is
responsible for the vanishing of the spectral function $\A_{c}^{\rm
  eq} (\omega)$ for $\omega \to 0$. Since we have derivative for
both $\hat{c}$ and $\hat{c}^{\dagger}$ in the Green function, and each
gives an extra factor of $\omega$, we end up with $\A_{c}^{\rm eq}
(\omega) \sim \omega^2$. This behaviour depends on being at half
filling (particle-hole symmetry), hence is modified when $\epsd$ is
not zero.

Now we can discuss the higher spectral functions, $\A_{d
  c^\dagger}^{\rm eq} (\omega)$, and $\A_{d c}^{\rm eq}
(\omega)$. These are the imaginary parts of the corresponding retarded
Green functions, up to a factor of $-1/\pi$. The retarded Green
functions are in turn the analytical continuation of the temperature
Green functions to the real frequency axis. And the temperature Green
functions can be found in the noninteracting case using Wick's
theorem. \cite{Mahan}

Performing these calculations for $\A_{d c^\dagger}^{\rm eq} (\omega)$, one gets: 
\begin{eqnarray}
	\A_{d c^\dagger}^{\rm eq} (\omega) & = & \frac{\rho}{\pi} \Im
	\left[ \ln \frac{\omega - \epsd + i\Gamma}{-\epsd + i\Gamma} \right.
	\\
	& & \left. -\frac{\Gamma^2}{\omega(\omega+2i\Gamma)}
	\ln \frac{\epsd^2 - ( \omega + i\Gamma )^2}{\epsd^2 + \Gamma^2} \right]
	\punkt \nonumber
\end{eqnarray}
Concentrating on $\omega \ll \Gamma$ one finds $\eta_{d
  c^\dagger}^{\rm eq} = \eta_{c}^{\rm eq} + \eta_{d}^{\rm eq} = 1$ for
$\epsd \ne 0$ and $\eta_{d c^\dagger}^{\rm eq} = \eta_{c}^{\rm eq} +
\eta_{d}^{\rm eq} = 2$ for $\epsd = 0$ [the data in
\frefsub{fig:AOC02spec}{c} corresponds to the latter case, which 
reproduces \Eq{eq:SIRL_B_FL-cddagger}]. This simple summation of
scaling dimensions is natural here, since there is only one possible
different-time Wick-pairing, of each single-particle operator with its
conjugate.

For $\A_{d c}^{\rm eq} (\omega)$, however, there are two different-time Wick-pairings, causing 
cancellations, and resulting in:
\begin{eqnarray}
	\A_{d c}^{\rm eq} (\omega) = \frac{\rho}{\pi} \Im
	\left[ \frac{\omega - 2 \epsd + 2 i \Gamma}{\omega - 2 \epsd}
	\ln \frac{\omega - \epsd + i\Gamma}{-\epsd + i\Gamma} \right]
	\punkt
\end{eqnarray}
Concentrating again on $\omega \ll \Gamma$ one finds now that 
$\eta_{d c}^{\rm eq} = 2$ for \emph{all} values of $\epsd$ [which 
reproduces \eref{eq:SIRL_B_FL-cddagger}].\cite{Goldstein2010b} The
reason is that in the low-energy continuum limit the product $\hat{d}
\hat{c}$ becomes the product of annihilation operators at almost the
same point. Hence, one should expand in the distance between $\hat{d}$
and $\hat{c}$ (lattice spacing). The leading term (with no spatial
derivatives) vanishes by Pauli's principle; the next term involves a
spatial derivative, leading to a factor of $\omega$, similarly to the
arguments above. Another factor of $\omega$ comes from the operator
$\hat{c}^{\dagger} \hat{d}^{\dagger}$ appearing in the definition of
$\G_{d c}^{\rm eq} (\omega)$. Thus, at low energies we end up with
$\A_{d c}^{\rm eq} (\omega) \sim \omega^3$ even for $\epsd \ne 0$.

Although the above calculations were performed for the noninteracting 
case, the qualitative arguments explaining the low-energy behaviour are 
valid even when $U > 0$. Moreover, since the system flows to the same 
fixed point for all values of $U > 0$, the low energy power-laws are in 
any case \emph{independent of $U$}. Our numerical results (\fref{fig:AOC02spec}) 
are in agreement with this picture.

Let us now turn to the low-energy behaviour of the PS setup in the 
case where PS is continuous. At low-energy the system is governed 
by Kondo physics, \eref{eq:KondoHamilton}, \cite{Kashcheyevs2007,
Lee2007,Silvestrov2007,Kashcheyevs2009} where the L-R degree of 
freedom plays the role of a pseudo-spin.  The equivalence to Kondo 
continues to hold even in the presence of a charge sensor, as shown 
by GBG. \cite{Goldstein2010,Goldstein2011} The operator $\hat{Y} 
= \hat{d}_{R}^{\dagger} \hat{d}_{L}^{\pdag} \hat{c}_{R}^{\pdag} 
\hat{c}_{L}^{\dagger}$ (which is relevant in the continuous-PS phase) 
is the pseudospin-flip local exchange term between the dot and the 
lead. Similarly, the IRLM is also equivalent to the Kondo model, \cite{
Gogolin,Schlottmann1980,Borda2008} with the role of spin replaced 
by the charging state of the dot. The pseudo-spin local exchange term 
is simply $\hat{d} \hat{c}^{\dagger}$. Hence, when the parameters are 
properly mapped, the spectral functions $\A_{Y}^{\rm eq} (\omega)$ 
and $\A_{d c^\dagger}^{\rm eq} (\omega)$ are equivalent when the 
Kondo description applies (i.e.\ for $\omega \ll D$ for the IRLM, and 
$\omega \ll \omega_{\rm he}$ for the PS setup). In particular, 
$\A_{Y}^{\rm eq} (\omega)$ should exhibit an $\omega^3$ behaviour 
at low energy, similarly to $A_{d c^\dagger}^{\rm eq} (\omega)$ for 
$\epsd = 0$, as the NRG data shows [dotted line in \frefsub{fig:PS_A_Spectra}{a} 
and \frefsub{fig:PS_C_Spectra}{a}].

%---------------------------------------------------------------------

\end{document}